\documentclass[%
 reprint,
 amsmath,amssymb,
 aps,
]{revtex4-2}

\usepackage{graphicx}
\usepackage{dcolumn}
\usepackage{bm}
\usepackage{xcolor}
\usepackage{multirow}
\usepackage{soul}

\begin{document}

\preprint{APS/123-QED}


\title{Active particles in power-law
potentials: steady state distributions and shape transitions}

\author{Abhik Samui$^{\dag}$}
\author{Manoj Gopalakrishnan$^{\dag\ddag}$}%
 \email{Corresponding author : manojgopal@iitpkd.ac.in}
\affiliation{%
 $^{\dag}$Department of Physics, Indian Institute of Technology Madras, Chennai 600036, India 
}%
\affiliation{%
 $^{\ddag}$Department of Physics, Indian Institute of Technology Palakkad, Palakkad 678623, India 
}%




\date{\today}

\begin{abstract}
We study the stationary states of an active Brownian particle (ABP)  and run-and-tumble particle (RTP) in  two dimensional power-law potentials, in the limit where translational diffusion is negligible. The potential energy of the particle has the form $U(r)\propto r^{n}$, where $n\geq 2$ and even. In two dimensions, we derive the exact equations for the positional probability distribution $\phi({\bf r})$ of ABP ($n\geq 2)$ and RTP ($n=2$), whose solutions are obtained under the assumption that the particle's orientation angle is Gaussian. Both analytical and numerical results show that, in all cases,  $\phi({\bf r})$ has compact support and undergoes a phase transition-like shape change as a function of the trap strength. For ABP, our theory predicts a continuous transition in shape for $n=2$ and a discontinuous transition for $n>2$, both of which agree with the simulation results. Simulations suggest the existence of both types of shape transition in the case of RTP as well. For ABP, in the strongly active regime, the orientational probability distribution is unimodal near the outer boundary but becomes bimodal towards the interior, signifying a transition from predominantly radial orientation to orbiting motion. In RTP, the analogous shape transition in the orientational distribution is almost absent. 
\end{abstract}

\maketitle



\section{Introduction}\label{introduction}
In recent years, the study of active matter has emerged as a hot topic of research, mainly due to the many associated collective phenomena, e.g., pattern formation~\cite{surrey2001physical}, active turbulence~\cite{dombrowski2004self,thampi2016active}, flocking~\cite{toner2005hydrodynamics,kumar2014flocking}, motility induced phase separation~\cite{schwarz2012phase,redner2013structure,stenhammar2015activity}, clustering~\cite{fily2012athermal,palacci2013living} and active pressure without equation of state~\cite{solon2015pressure}. Active systems are internally driven out of equilibrium, where each constituent particle converts the energy extracted from its local environment into systematic movements~\cite{schweitzer2003brownian}. 
In nature, they span a wide range of length and energy scales, from single cells such as bacteria~\cite{berg2004coli,sokolov2012physical} and spermatozoa~\cite{riedel2005self}, intracellular components such as microtubules~\cite{sanchez2012spontaneous}, actin filaments ~\cite{mizuno2007nonequilibrium} and motor proteins~\cite{carter2005mechanics}, to macroscopic entities such as flock of birds, school of fish or herd of animals~\cite{ramaswamy2010mechanics,vicsek2012collective,marchetti2013hydrodynamics}. In addition, active particles can also be inanimate objects such as vibrated granular rods~\cite{narayan2007long,kudrolli2008swarming}, and phoretic colloids~\cite{paxton2004catalytic,howse2007self}. 

Away from the realm of collective dynamics, a single active particle may also exhibit highly non-intuitive behavior compared to its passive counterpart, such as accumulation near a physical boundary~\cite{li2009accumulation,elgeti2015run,wagner2017steady} and non-Boltzmann distributions inside a potential well~\cite{erdmann2000brownian,takatori2016acoustic,das2018confined,buttinoni2022active}. In recent experiments, the catalytic Janus particle~\cite{howse2007self} has been widely used to explore various aspects of active dynamics~\cite{takatori2016acoustic,li2017two}. Takatori et al.~\cite{takatori2016acoustic} showed that non-interacting Janus particles inside an acoustic trap give rise to a non-Boltzmann positional distribution in a stronger trap, whereas for a weaker trap, the distribution remains Boltzmann-like.
In contrast, Buttinoni et al~\cite{buttinoni2022active} found a non-Boltzmann positional distribution for an active Janus particle inside an optical trap when the trap strength is weak. Schmidt et al. demonstrated that an active nanoparticle trapped by an optical tweezer is characterized by non-Boltzmann positional distribution and orbital rotation beyond a critical temperature~\cite{schmidt2021non}. Remarkably, even a macroscopic robot moving on a parabolic disk was shown to exhibit highly non-intuitive behaviors such as climbing and orbiting motions~\cite{dauchot2019dynamics}.    

The non-trivial and counter-intuitive nature of the stationary state of an active particle in confining potentials has been the subject of many theoretical studies~\cite{vladescu2014filling,  hennes2014self, elgeti2015run,  malakar2018steady, duzgun2018active, basu2019long, malakar2020steady, chaudhuri2021active, wagner2022steady,  nakul2023stationary, knippenberg2024motility,santra2021direction}. Simple stochastic models such as Active Brownian Particle (ABP) and Run and Tumble Particle (RTP), as well as their variants, have been of great help in obtaining valuable insights into the dynamics and stationary states of active particles.
The ABP, in particular, has become a paradigmatic model for active transport~\cite{schimansky1995structure,romanczuk2012active}. The dynamics of an ABP consists of deterministic self-propulsion as well as continuous stochastic reorientation of the direction of self-propulsion by rotational diffusion. In addition, the particle may also be subjected to random translational diffusion. Indeed, in the limit of vanishing active velocity, the ABP behaves like a passive Brownian particle~\cite{romanczuk2012active,nakul2023stationary}. On the other hand, the dynamics of a RTP~\cite{berg2004coli,tailleur2008statistical,santra2020run} consists of, in addition to the features mentioned above, discontinuous changes in its  orientation at randomly chosen instants, such that the duration between two consecutive tumbles is specified by a Poisson distribution~\cite{berg1972chemotaxis}. 

From a number of experimental, computational and theoretical studies, it is now well-known that in the presence of a confining potential, ABP tends to stick close to the ``outer boundary"~\cite{rothschild1963non,berke2008hydrodynamic,yang2014aggregation,takatori2016acoustic,nakul2023stationary,malakar2020steady}. In soft, symmetric potentials, this outer boundary is a ring-like region around the center of the potential where the mean velocity vanishes. This non-trivial, ``anti-Boltzmann" stationary state is realized when (a) the confining potential is sufficiently steep and/or  (b) the persistence length of active motion is large, corresponding to large active velocity and/or small rotational diffusivity~\cite{malakar2020steady,basu2019long,nakul2023stationary}. This can be understood as follows: (a) If the active velocity is very small, ABP can be approximated as a Brownian particle, any movement would be due to the diffusion. As the particle is confined under an axially symmetric potential, eventually a steady state will be reached, in which the particle's positional probability distribution resembles the Boltzmann distribution with a peak at the center~\cite{buttinoni2022active}. In contrast, (b) when the ABP has a sufficiently large active velocity, self-propulsion will dominate over translational diffusion. A fast-moving ABP (higher persistence length) with a significant radial component of velocity will bounce off the outer boundary, and taking advantage of its large active velocity, will immediately reach another point on the boundary where it goes through the process again. If the tangential component of the velocity is dominant, the particle will be found to be sliding along the boundary~\cite{dauchot2019dynamics,vincent2025curvature} for long intervals of time. As a result, on the average, the particle spends a large fraction of the time close to the boundary with a corresponding maximum in the positional probability distribution~\cite{malakar2020steady,basu2019long,nakul2023stationary}. By the same reasoning, in this regime, the angular distribution becomes bimodal near the wall, with two symmetrically located peaks corresponding to nearly clockwise and nearly anti-clockwise motion~\cite{dauchot2019dynamics,nakul2023stationary}.

To theoretically understand the dynamics of an active particle inside a potential well, a simplest and convenient choice of potential would be a harmonic trap. Even for a harmonic trap, it is generally difficult to solve the Fokker-Planck equation for the complete probability distribution exactly. Malakar et al.~\cite{malakar2020steady} obtained the power series solution to the Fokker-Planck equation of an ABP confined in a two dimensional harmonic trap and confirmed the passive to active transition observed in experiments~\cite{takatori2016acoustic}.
They also predicted a remarkable re-entrant active to passive transition at very high stiffness of the potential. Basu et al.~\cite{basu2019long} found an exact recursion relation for the time-dependent moments of an ABP in a two dimensional harmonic trap when translational diffusion is absent. Chaudhuri and Dhar\cite{chaudhuri2021active} used a Laplace transform-based method to find all relevant moments analytically as a function of time for an ABP confined in a harmonic well in $d$ dimensions. The authors analyzed the kurtosis of the position variable and found the active-passive cross-over regions in $d$ dimensions. More recently, Nakul and Gopalakrishnan~\cite{nakul2023stationary} computed the steady state positional distribution for an ABP in a two dimensional harmonic trap, under the assumption that the orientation angle is Gaussian. They also showed the possible existence of a continuous shape transition of the positional distribution as a function of activity.  

In the case of RTP, in $d=1$, Malakar et al.~\cite{malakar2018steady} found the exact positional distribution for both bounded and unbounded regions by solving the generalized Telegrapher's equations. They showed that, in steady state, the positional distribution in the unbounded region approaches a Gaussian distribution, whereas for the bounded region, the distribution has peaks near the boundary in steady state. Basu et al.~\cite{basu2020exact} solved a model for RTP in a harmonic potential in one dimension where RTP has three internal states, and found its exact steady state positional distribution . This work was extended by Smith et al.~\cite{smith2022exact} to a two-dimensional RTP in a harmonic potential, and the time-dependent positional distribution and the corresponding steady state expression were found exactly: in their model, the RTP changes its orientation discretely by an angle $\pi/2$. In another work for RTP in a two dimensional harmonic potential, Frydel et al.~\cite{frydel2022positing} obtained an exact linear differential equation of the positional distribution in steady state and found its solution as the convolution of two independent distributions. 

The dynamics of active particles in more general confining potentials has also been studied. Dhar et al.~\cite{dhar2019run} studied one-dimensional motion of a RTP in a potential of the form $U(x)=\alpha|x|^p$ where $\alpha, p>0$. They described the steady state active-passive crossover regions in the ($p$,$\alpha$) plane in detail and found the exact analytical results for the relaxation and first passage process for $p=1$ and $p=2$. Sevilla et al.~\cite{sevilla2019stationary}, studied the superstatistics distributions of run and tumble particles in steady state in various one-dimensional confining potentials,including harmonic trap. 

Despite such numerous studies, an understanding of the stationary states of ABP confined in super-harmonic power-law-type potential wells in two and higher dimensions is still largely absent, unlike in one dimension~\cite{sevilla2019stationary,dhar2019run}. A vast majority of the existing literature focuses on the positional distribution of the ABP, and less on the orientational distribution, although, as has been pointed out in previous studies~\cite{buttinoni2022active, nakul2023stationary}, the latter has many interesting features of its own. In this article, we try to address these gaps. Here, we study the general stationary states of active particles, both ABP and RTP, in power-law potential wells of the general form $U(r)\propto r^n$ where $n\geq2$ and even, in spatial dimension $d=2$. In the case of ABP, starting from the general Fokker-Planck equation, we obtain a closed expression for the positional distribution for arbitrary values of $n$. In addition, we also make considerable analytical progress towards understanding the interesting and non-trivial features of the angular distribution. For RTP, using a similar method, we find a closed form solution for the positional distribution in the harmonic potential, and show that it undergoes a continuous shape transition as a function of the trap strength, like the ABP~\cite{nakul2023stationary}. For $n>2$, the shape transition becomes discontinuous for both ABP and RTP, as observed in our numerical simulations. In the case of ABP, the numerical results are completely in agreement with our theoretical predictions. 

The structure of this paper is as follows. We describe the essential features of the model in Section~\ref{model}. The Fokker-Planck equation-based formalism for the ABP is developed, its solutions are derived and compared against numerical simulation data in Section~\ref{Fokker_Planck_equation_solution}.  In Section~\ref{sec:RTP}, we extend our formalism to RTP and investigate its steady states in two different power-law potentials, harmonic and super-harmonic. In Section~\ref{conclusions}, we summarize our important results with a brief discussion and outline our conclusions. 
\section{Model Details}\label{model}

We visualize a model ABP as a spherical particle, but with an entrenched asymmetry between its two hemispheres. The symmetry axis runs from the north pole to the south pole and defines the direction of active propulsion. We denote the self-propulsion velocity by ${\bf u}=u_0\hat{\mathbf{u}}$. The unit vector $\hat{\mathbf{u}}$ that specifies the direction of self-propulsion is a random variable, subject to continuous Brownian rotation. Let ${\bf r}$(t) denote the position of the particle.

In the overdamped regime, the dynamics of the ABP in the presence of a (spherically symmetric) force-field with potential energy $U(r)$ is described by the Langevin equations
\begin{equation}\label{eq:overdamped_langvin_eq}
\begin{aligned}
    \frac{d\mathbf{r}(t)}{dt} = u_{0} \, \hat{\mathbf{u}}(t) - \frac{U^{\prime}(r)}{\gamma_t}\hat{\mathbf{r}}  + \sqrt{2D_t} \, {\boldsymbol{\eta}_t}(t) \\[12pt]
    \frac{d\hat{\mathbf{u}}(t)}{dt} = \sqrt{2 D_r}\, {\boldsymbol{\eta}_r}(t) \times \hat{\mathbf{u}}(t),
\end{aligned}
\end{equation}
where $D_t=k_B T \,\gamma_t^{-1}$ and $D_r=k_B T \,\gamma_r^{-1}$ are the translational and rotational diffusion coefficients respectively, with $\gamma_t^{-1}$ and $\gamma_r^{-1}$ being the corresponding mobilities. For a spherical particle of radius $a$, $\gamma_t=6\pi\eta a$ and $\gamma_r=8\pi\eta a^3$, where $\eta$ is the dynamic viscosity of the fluid. In Eq.~\ref{eq:overdamped_langvin_eq}, $\boldsymbol{\eta_t}$ and $\boldsymbol{\eta_r}$ are the Gaussian white noises for translational and rotational motion, respectively. Both noise terms have zero mean and are uncorrelated in time: $\langle \boldsymbol{\eta}_t(t) \cdot  \boldsymbol{\eta}_t(t^{\prime}) \rangle \,= d \,\delta(t - t^{\prime})$ and $ \langle \boldsymbol{\eta}_r(t) \cdot  \boldsymbol{\eta}_r(t^{\prime}) \rangle \,= (d-1)\, \delta(t - t^{\prime})$, where $d$ is the dimension of the spatial region where the particle is moving. 

We choose the potential energy to be power-law type: 
\begin{equation}
    U(r) = \frac{\epsilon_0}{n} \left( \frac{r}{R} \right) ^{n}
    \label{eq2}
\end{equation}
where $\epsilon_0>0 $ has the dimension of energy, $r$ is the radial distance from the origin and $R$ is an arbitrary length scale. It is convenient to choose 
\begin{equation}
R=\frac{\epsilon_0}{\gamma_t u_0}
\label{eq3}
\end{equation}
whose significance will become clear soon. Henceforth, we refer to this length scale as the {\it trap radius} for the given potential. 

\begin{figure}[]
\centering
  \includegraphics[scale=0.30]{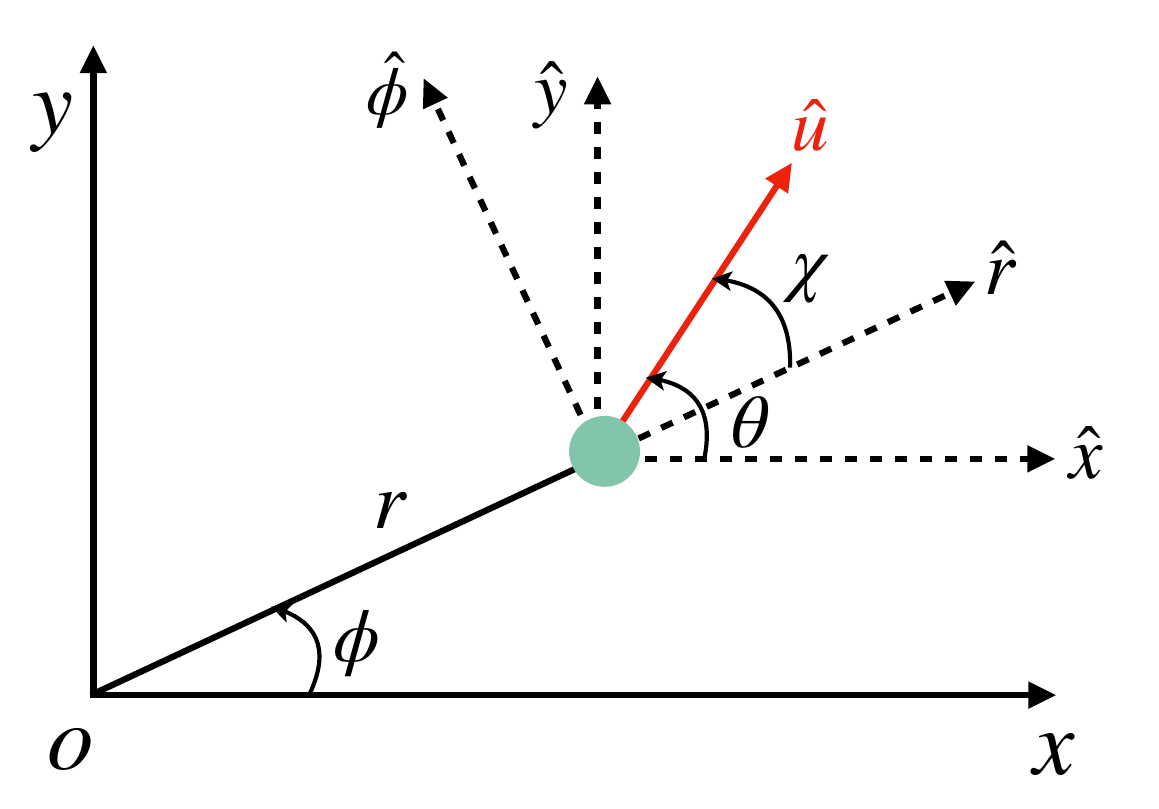}
  \caption{Schematic figure showing the relevant dynamical variables of an active particle in two dimensions. Here, ($r, \, \phi$) are the polar coordinates for the particle's position while $\hat{\mathbf{u}}$ is a unit vector indicating the direction of self-propulsion.}
  \label{fig:schematic_abp_rtp}
\end{figure}

In $d=2$, the unit orientation vector is given by $\hat{\mathbf{u}}=\cos\theta\, {\hat{\mathbf{x}}}+ \sin\theta\, {\hat{\mathbf{y}}}$, 
with ${\hat{\mathbf{x}}}$ and ${\hat{\mathbf{y}}}$ being the unit Cartesian vectors along the $x$ and $y$ axes respectively (Fig.~\ref{fig:schematic_abp_rtp}). In this case, the Langevin equation for the orientation angle is simplified to 
\begin{equation}
\frac{d\theta}{dt}=\sqrt{2D_r}\,\eta_r(t)
\label{eq4}
\end{equation}
with $\langle \eta_r(t)\rangle=0$ and $\langle\eta_r(t)\eta_r(t^{\prime})\rangle=\delta(t-t^{\prime})$. 
In RTP, the orientational dynamics involves, in addition to the continuous Brownian rotation above, discontinuous changes (tumbles), where during a small time interval $dt$, $\theta\to \theta^{\prime}$ with probability $\lambda dt$, where $\lambda$ is the Poisson rate for the tumble process. 

\section{ABP in power-law potential well}\label{Fokker_Planck_equation_solution}

\subsection{The stationary Fokker-Planck equation}
Let $P(\mathbf{r},\hat{\mathbf{u}},t)$ be the joint probability density for the position ${\bf r}$ and the orientation $\hat{\mathbf{ u}}$ of the ABP in the potential well. For a free ABP in two dimensions, the Fokker-Planck equation has been derived and studied in \cite{sevilla2015smoluchowski}. In arbitrary spatial dimension $d$, and in the presence of a potential well, the Fokker-Planck equation can be written as 
\begin{equation}
    \frac{\partial P(\mathbf{r},\hat{\mathbf{u}},t)}{\partial t} = - \boldsymbol{\nabla} \cdot \mathbf{J}_t + D_r {\nabla}^{2}_{\hat{\mathbf{u}}} P(\mathbf{r},\hat{\mathbf{u}},t)
    \label{FPE}
\end{equation}
where $\mathbf{J}_t$ is the probability current density vector for translational motion and ${\nabla}^{2}_{\hat{\mathbf{u}}}$ is the angular Laplacian. The general expression of $\mathbf{J}_t$ is given by 
\begin{equation}
    \mathbf{J}_t = -D_t \boldsymbol{\nabla} P + \mathbf{V} P
    \label{eq:prob_current_density}
\end{equation}
where
\begin{equation}\label{eq:deterministic_vel}
    \mathbf{V} = u_0\Bigl[\hat{\mathbf{u}} - \left( \frac{r}{R} \right) ^{n-1} \hat{\mathbf{r}} \Bigr]
\end{equation}
is the deterministic part of the translational velocity of ABP in power-law potential well. In stationary state, the time derivative becomes zero, hence the stationary Fokker-Planck equation (SFPE) is given by 
\begin{equation}\label{eq:SFPE}
      - \boldsymbol{\nabla} \cdot \mathbf{J}_t + D_r {\nabla}^{2}_{\hat{\mathbf{u}}} P(\mathbf{r},\hat{\mathbf{u}}) =0.
\end{equation}
To solve Eq.~\ref{eq:SFPE}, we follow the same methodology as used in \cite{nakul2023stationary}. The stationary state probability distribution $P(\mathbf{r},\hat{\mathbf{u}})$ can be written as a product of the purely positional probability density and a conditional angular probability distribution: 
\begin{equation}
    P(\mathbf{r}, \hat{\mathbf{u}})= \Phi(\mathbf{r}) f(\hat{\mathbf{u}}|\mathbf{r})
    \label{eq9}
\end{equation}
where
\begin{equation}
    \Phi(\mathbf{r}) = \int P(\mathbf{r},\hat{\mathbf{u}}) \, d \hat{\mathbf{u}}.
    \label{eq10}
\end{equation}
It is useful to define the conditional average of the unit propulsion vector: 
\begin{equation}
    \Bar{\hat{\mathbf{u}}}({\bf r}) = \int \hat{\mathbf{u}} \, f(\hat{\mathbf{u}}|\mathbf{r}) \, d\hat{\mathbf{u}}
    \label{eq11}
\end{equation}
By integrating Eq.~\ref{eq:SFPE} with respect to $\hat{\mathbf{u}}$, we arrive at the simple divergence-free condition 
\begin{equation}\label{eq12}
    \boldsymbol{\nabla} \cdot \mathbf{K}(\mathbf{r}) = 0
\end{equation}
for the effective translational current density $\mathbf{K}(\mathbf{r}) = \int \mathbf{J}_t \, d\hat{\mathbf{u}} \,$, which is given by the explicit expression
\begin{equation}\label{eq:expression_of_K}
    \mathbf{K}(\mathbf{r}) = -D_t \boldsymbol{\nabla}{\Phi(\mathbf{r})} + u_0\Bigl[\Bar{\hat{\mathbf{u}}} -  \left( \frac{r}{R} \right) ^{n-1} \hat{\mathbf{r}}  \Bigr] \, {\Phi}(\mathbf{r}).
\end{equation}
Due to the axial symmetry of the underlying potential, it follows that the tangential component $K_{\phi}=0$. Furthermore, in the absence of sources or sinks, we also expect the condition $K_r=0$ to hold, hence $\mathbf{K}(\mathbf{r})=0$ identically. The radial symmetry further imposes the following constraints on the positional probability density and the conditional orientation average:
\begin{equation}
    \Bar{\hat{\mathbf{u}}}(\mathbf{r}) = b(r) \hat{\mathbf{r}}~~;~~\Phi(\mathbf{r})=\Phi(r)
    \label{eq14}
\end{equation}
Define the dimensionless radial position coordinate $\xi = r/R$, the corresponding mean conditional orientation variable $g(\xi) \equiv b(R \, \xi)$ and the probability density $\Psi(\xi)=R^d \Phi(R \, \xi)$. From Eq.~\ref{eq:expression_of_K}, these quantities together satisfy the equation (after putting ${\bf K}=0$ as required by symmetry)
\begin{equation}\label{non_dimensional_K_zero_eq}
    -D^{\prime}_t \, \frac{\partial \Psi(\xi)}{\partial \xi} + [ g(\xi) -  \, \xi^{n-1}] \Psi(\xi) = 0
\end{equation}
where 
\begin{equation}
D^{\prime}_t = \frac{D_t\gamma_t}{\epsilon_0}
\label{eq16}
\end{equation}
is a dimensionless translational diffusion coefficient (inverse P\'eclet number). For the rest of this paper, we will focus on the regime $D_t^{\prime}\to 0$, and put $D_t=0$ in all our equations and simulations. In this limit, from Eq.~\ref{non_dimensional_K_zero_eq}, we find that 
\begin{equation}\label{eq:g_of_chi}
    g(\xi)\sim \xi^{n-1}
\end{equation}
indicating that the radial orientation becomes more prominent as $\xi$ increases. 
\begin{figure*}[]
\centering
  \includegraphics[scale=0.45]{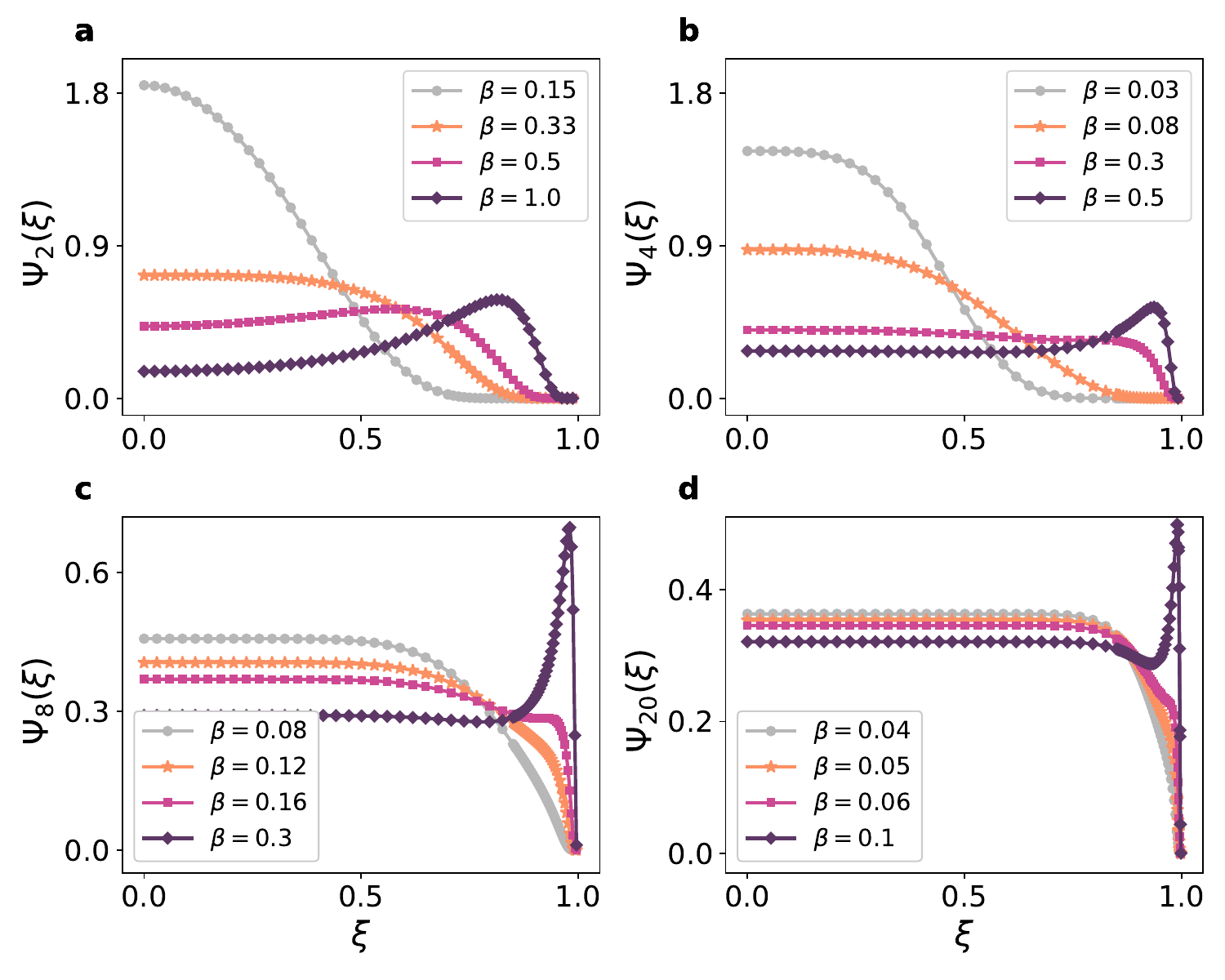}
  \caption{The positional distribution $\Psi_n(\xi)$ for the ABP, predicted by theory (Eq.~\ref{expression_of_psi}) is plotted against $\xi$ for different values of $\beta$ and $n$, as indicated. To ensure series convergence, the upper limit of the summation over $s$ in Eq.~\ref{expression_of_psi} is set as 500. In each figure, $n$ is fixed (as $2,4,8$ and $20$ for $\textbf{a},\textbf{b}, \textbf{c}$ and $\textbf{d}$ panels respectively), and the  different colors (or shapes) correspond to different values of $\beta$.}
  \label{fig:Psi_at_diif_beta_for_n}
\end{figure*}

\subsection{The equation for $\Psi(\xi)$}
An equation for $\Phi(r)$ can be derived by multiplying Eq.~\ref{eq:SFPE} by $\mathbf{V}$, and then integrating it  over $\hat{\mathbf{u}}$. The result turns out to be (see Appendix~\ref{Integration_I_1_and_I_2} for details)
\begin{equation}\label{FPE_in_d_dimension}
    \begin{split}
  &  u_0^2 \int \hat{\mathbf{u}} (\hat{\mathbf{u}} \cdot \boldsymbol{\nabla} P) \, d \hat{\mathbf{u}} \, - \frac{u_0^2}{R^{2(n-1)}} \, (2n -3 +d)\, \Phi\, r^{2n-4} \, \mathbf{r}\, \\
  & - \frac{u_0^2}{R^{2(n-1)}} r^{2n-4} \mathbf{r} \, (\mathbf{r} \cdot \boldsymbol{\nabla} \Phi)\\
  & - u_0\, D_r (1-d) \, \Phi \, \frac{1}{R^{n-1}}\, r^{n-2} {\mathbf{r}} =0
    \end{split}
\end{equation}
where we have used the definition in Eq.~\ref{eq3}. 

The rest of the paper is restricted to $d=2$. In terms of the scaled coordinate $\xi$, the radial part of Eq.~\ref{FPE_in_d_dimension} is reduced to 

\begin{equation}\label{FP1}
\begin{split}
  &  \hat{\mathbf{r}} \cdot \int \hat{\mathbf{u}} \, (\hat{\mathbf{u}} \cdot \boldsymbol{\nabla}_{\xi} \mathcal{P}) \, d\theta \\
  & = (2n-1)\, \xi^{2n-3}\, \Psi +\, \xi^{2n-2} \,\frac{\partial \Psi}{\partial \xi} -\frac{1}{\beta}\, \xi^{n-1}\Psi
\end{split}
\end{equation}
where 
\begin{equation}
\beta= \frac{u_0}{R D_r}
\label{eq20}
\end{equation}
is dimensionless activity, ${\boldsymbol\nabla}_{\xi}\equiv R\boldsymbol{\nabla}$ and $\mathcal{P}({\boldsymbol\xi},\theta)=R^2 P({\bf r},\theta)$ where ${\bf r}=(r,\phi)$. Radial symmetry implies that $\mathcal{P}({\boldsymbol \xi}, \theta) \to \mathcal{P}(\xi, \chi) = \Psi(\xi) \, f(\chi | \xi) $ where $\chi=\theta - \phi$~\cite{malakar2018steady} and $f(\chi | \xi)$ is the conditional probability density of the same.  The l.h.s in Eq.~\ref{FP1} is evaluated in Appendix~\ref{calculation_of_u_u_gradP_int}, with the result, 
\begin{equation}\label{FP2}
\begin{split}
  &  \hat{\mathbf{r}} \cdot \int \hat{\mathbf{u}} \, (\hat{\mathbf{u}} \cdot \boldsymbol{\nabla}_{\xi} \mathcal{P}) \, d\theta\ \\
  & = \Bigl[\sigma_{\cos \chi}^2 + \xi^{2n-2}  \Bigr] \partial_{\xi} \Psi \\
  & + \Psi \Bigl[\partial_{\xi} \sigma_{\cos \chi}^2 + \frac{2 \sigma_{\cos \chi}^2}{\xi} - \frac{1}{\xi} +2n\xi^{2n-3} \Bigr]. 
\end{split}
\end{equation}
where $\sigma^2_{\cos\chi}(\xi)$ is the conditional variance of $\cos\chi$, evaluated using the conditional probability distribution function $f(\chi | \xi)$. 

After substituting Eq.~\ref{FP2} in Eq.~\ref{FP1}, we arrive at  the exact equation for $\Psi(\xi)$: 
\begin{equation}\label{FP3}
\begin{split}
  &  \sigma_{\cos \chi}^2  \frac{\partial \Psi}{\partial \xi}+\Bigl[\partial_{\xi} \sigma_{\cos \chi}^2 + \frac{2 \sigma_{\cos \chi}^2}{\xi} - \frac{1}{\xi} + \xi^{2n-3} + \frac{1}{\beta} \xi^{n-1} \Bigr]\Psi, \\
  & =0.
\end{split}
\end{equation}
In order to make further progress, it is useful to assume that $f(\chi | \xi)$ is Gaussian, in which case it can be shown that ~\cite{nakul2023stationary} $\sigma_{\cos \chi}^2 = \frac{1}{2} (1 - \langle\cos \chi \rangle^2)^2$. Using the expression in Eq.~\ref{eq:g_of_chi} for the mean, it then follows that   
\begin{equation}
\sigma_{\cos \chi}^2\simeq \frac{1}{2} (1 - \xi^{2(n-1)})^2.
\label{eq23}
\end{equation}
Using Eq.~\ref{eq23} in Eq.~\ref{FP3}, we arrive at the following approximate equation for $\Psi(\xi)$:
\begin{equation}\label{diffential_eq_FP}
    \begin{split}
     &  \frac{1}{2} \bigg[1 - \xi^{2(n-1)}\bigg]^2 \frac{\partial \Psi}{\partial \xi} \\
     & -\Bigl[(2n-1) \xi^{2n-3}  \bigl\{1 - \xi^{2n-2}  \bigr\} - \frac{1}{\beta} \xi^{n-1}   \Bigr]\Psi=0
    \end{split}
\end{equation}
whose solution (for details, see Appendix~\ref{simplification_int_dPsi_by_Psi}) is given by 
\begin{equation}\label{expression_of_psi}
    \begin{split}
          \Psi_n(\xi) = & \,C_{n} \Bigl[1 - \xi^{2(n-1)} \Bigr]^{-\frac{(2n-1)}{(n-1)}} \\
          & . \exp \Bigl[-\frac{2\xi^n}{n \beta} \Bigl\{ 1 + \sum_{s=1}^{\infty} \frac{(s+1)(\frac{n}{n-1})}{(2s + \frac{n}{n-1})}\, \xi^{2s(n-1)} \Bigr\}   \Bigr].
    \end{split}
\end{equation}
Note that the expression on the r.h.s is defined on the interval $[0:1]$. The integration constant $C_{n}$ is to be fixed by normalisation. In Appendix~\ref{aprrox_Psi_for_large_n}, we also derive a compact approximation (Eq.~\ref{closed_form_expression_large_n}) of Eq.~\ref{expression_of_psi} for large $n$, which turns out to be useful in comparisons with simulation data even for small and moderate values of $n$. 

For the special case $n=2$ (harmonic trap), Eq.~\ref{expression_of_psi} reduces to 
\begin{equation}\label{psi_harmonic_potential}
    \begin{split}
        & \Psi_2(\xi) = C_2\Bigl[1 -  \xi^{2} \Bigr]^{-3}\exp \Bigl[-\frac{ \xi^2}{\beta (1 -  \xi^2)} \Bigr], 
    \end{split}
\end{equation}
which matches the existing result derived in~\cite{nakul2023stationary}.


\begin{figure}[]
\centering
  \includegraphics[scale=0.35]{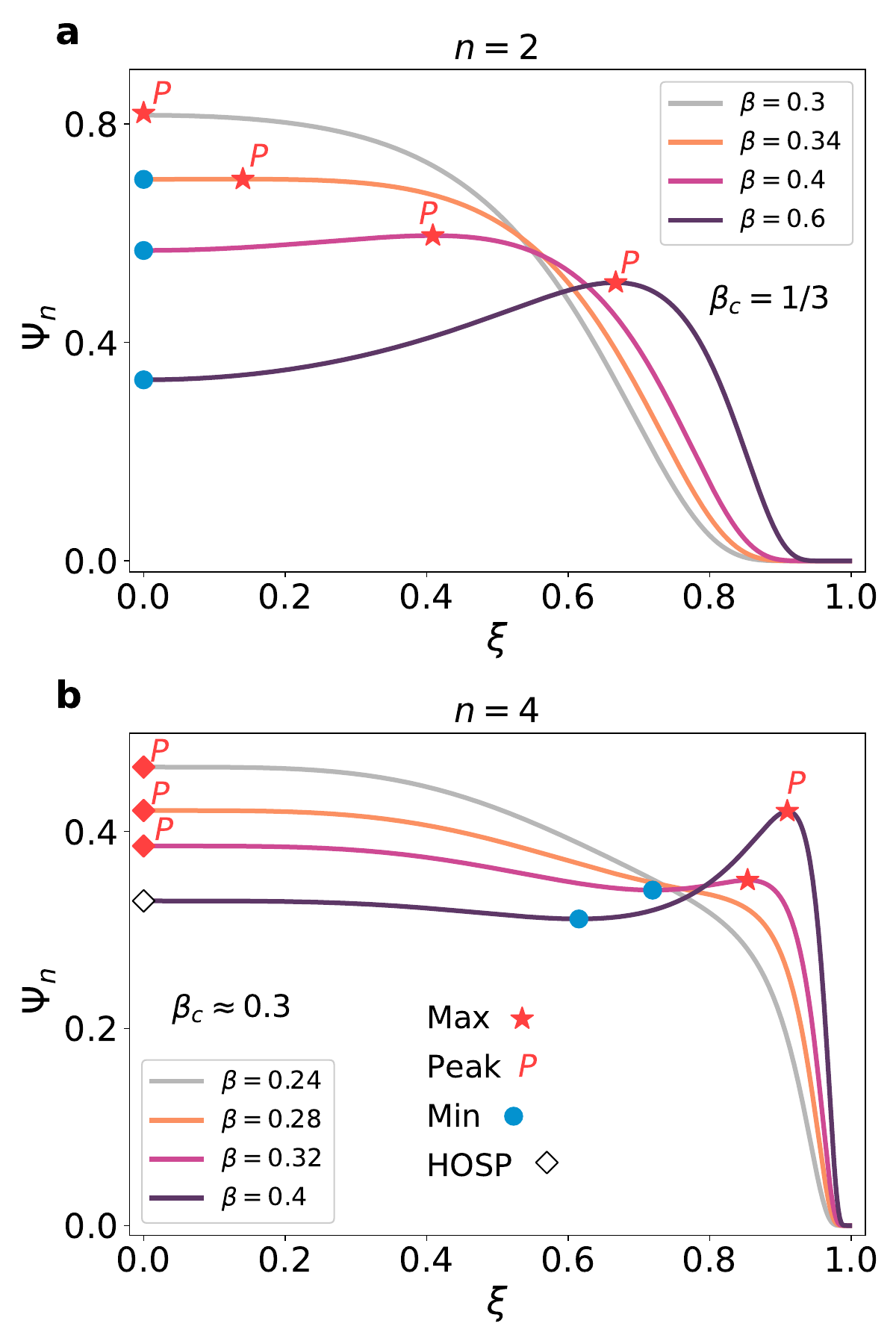}
  \caption{This figure demonstrates how the shape transition is different for $n=4$, from $n=2$. In both (a) and (b), the peak, denoted by the symbol red $P$ is the highest point of the distribution, while the red star (max) and blue circle (min) indicate local maxima and minima respectively. The diamond-shaped marker stands for a higher-order stationary point (HOSP), which is colored red if it corresponds to a peak.}
\label{fig:compare_peak_max_min_n_2_n_4}
\end{figure}

\begin{figure}[]
\centering
  \includegraphics[scale=0.35]{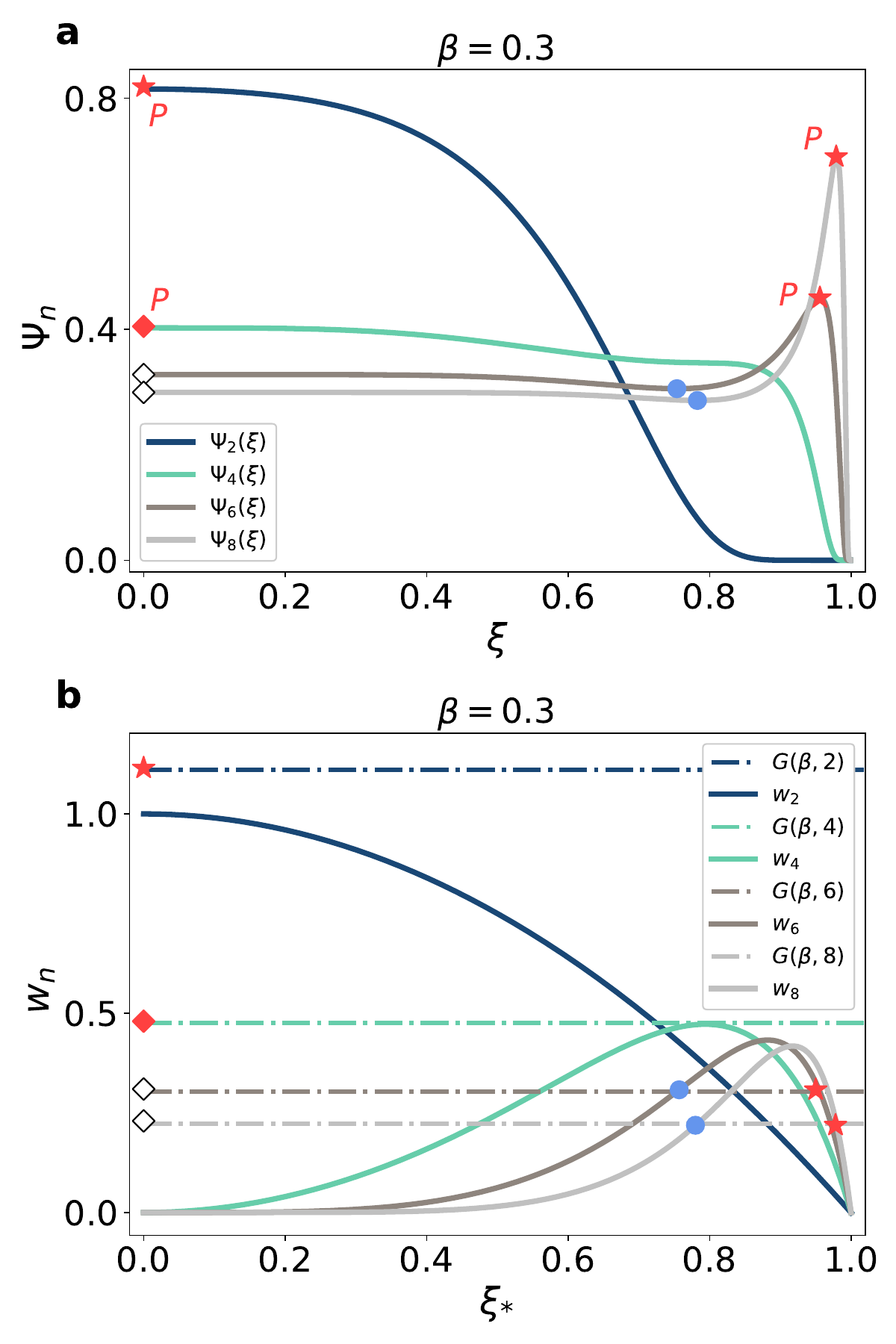}
  \caption{Shape transition in $\Psi_n(\xi)$ as a function of $n$ for a fixed $\beta$ is shown. Star (red) and circle (blue) shaped points indicate position of the maximum (or peak here) and minimum respectively, while the diamond shaped point represents a higher order stationary point (HOSP), which is colored red if it corresponds to a peak. The shape transition is discontinuous, as discussed in the text.}
\label{fig:xi_max_vs_n_ABP_for_constant_beta_0_3}
\end{figure}

\subsection{The nature of the positional distribution}
Fig.~\ref{fig:Psi_at_diif_beta_for_n} shows the distribution $\Psi_n(\xi)$ in Eq.~\ref{expression_of_psi} plotted against $\xi$, for various values of $\beta$ and $n$. In order to ensure convergence, the upper limit in the infinite summation was fixed at $500$. For fixed $n$ and increase in $\beta$, the peak of the distribution moves from $\xi=0$ to $\xi=1$. We show later, using rigorous arguments, that this change in the shape of the distribution $\Psi_n(\xi)$ is a {\it shape transition}, as has been established previously for $n=2$~\cite{nakul2023stationary}. In this case, theoretical arguments supported by numerical simulations have suggested a continuous shape transition where the peak remains at $\xi=0$ below a critical value of $\beta$, but continuously moves towards $\xi=1$ above it. For $n>2$, the corresponding observations (Fig.~\ref{fig:Psi_at_diif_beta_for_n} {\bf b},{\bf c},{\bf d}) are qualitatively similar, but the peak at $\xi=0$ at low $\beta$ is flatter when compared to $n=2$, and flattens even further as $\beta$ (or $n$) increases. For much larger $\beta$, a pair of maximum and minimum develop near the outer edge of the well, close to $\xi=1$. 

The occurrence of the pair of extrema close to $\xi=1$ with increase in $\beta$ and the transition in the shape of the distribution can be understood by studying the slope of $\Psi_n(\xi)$ (Eq.~\ref{diffential_eq_FP}). At the points of extremum, say, $\xi=\xi_*$, $h_{n}(\xi_*)=0$ where 
\begin{equation}
    h_{n}(\xi) = (2n-1) \Bigl \{\xi^{2n-3} - \xi^{4n-5}  \Bigr \} - \frac{1}{\beta} \xi^{n-1}. 
\end{equation}
Obviously, a trivial solution is $\xi_*=0$, while the non-trivial extrema satisfy the equation 
\begin{equation}
w_n(\xi_*)=G(\beta,n)
\label{eq28}
\end{equation}
where 
\begin{equation}
w_n(\xi)= \xi^{n-2}-\xi^{3n-4}~~;~~G(\beta,n)=[(2n-1)\beta]^{-1}
\label{eq29}
\end{equation}
For $n=2$, the above equation has a single solution,
\begin{equation}\label{eq30}
\xi_*=\sqrt{1-1/3\beta},
\end{equation}
provided $3\beta>1$. For $\beta<1/3$, the trivial solution $\xi_*=0$ is a maximum and the non-trivial solution does not exist. But when $\beta>1/3$, the non-trivial solution appears as a maximum and the trivial one, $\xi_*=0$ becomes a minimum.
Hence, $\beta_c(2)=1/3$ is the critical value of $\beta$ for $n=2$~\cite{nakul2023stationary}.

For $n>2$, the non-trivial extrema can be found graphically from Eq.~\ref{eq28} (for a specific example, see Fig.~\ref{fig:xi_max_vs_n_ABP_for_constant_beta_0_3}\textbf{b}). In the l.h.s of Eq.~\ref{eq28}, $w_n(\xi_*)$ is a non-monotonic function of $\xi_*$ in the range $[0:1]$. It vanishes at both $\xi_*=0$ and $\xi_*=1$, and has a single peak at $\xi_*=\Delta_n$, where
\begin{equation}
\Delta_n=\bigg(\frac{n-2}{3n-4}\bigg)^{\frac{1}{2(n-1)}}
\label{eq31}
\end{equation}
will be called the {\it gap parameter}, for reasons that will be clear soon. The maximum value of $w_n(\xi_*)$ is found by substituting Eq.~\ref{eq31} in Eq.~\ref{eq28} and Eq.~\ref{eq29}. Clearly, Eq.~\ref{eq28} has a single solution if $w_n(\Delta_n)= [(2n-1)\beta]^{-1}$ and two solutions (only one for $n=2$) if $w_n(\Delta_n)> [(2n-1)\beta]^{-1}$. We may thus identify a critical value for $\beta$, i.e., 
\begin{equation}
\beta_c(n)=[(2n-1)w_n(\Delta_n)]^{-1}
\label{eq32}
\end{equation}
such that, for $\beta>\beta_c(n)$, for $n>2$, the distribution $\Psi_n(\xi)$ will have two non-trivial extrema away from $\xi=0$, along with the trivial extremum $\xi_*=0$. For $\beta< \beta_c(n)$, however, $\xi_*=0$ is the only point of extremum. It can be shown that, for $n>2$, the trivial extremum is neither a maximum nor a minimum. Rather, since all derivatives $\partial^m \Psi_n/\partial \xi^m|_{\xi=0}=0$ for $m=1,2,...,n-1$, we refer to this extremum as a higher-order stationary point (HOSP). 

In the limit $n\to\infty$, it can be shown that $\Delta_n\sim 3^{-\frac{1}{2n}}$ and approaches 1 as $n\to\infty$. Substitution in Eq.~\ref{eq29} shows that, in this limit, $w_n(\Delta
_n)\sim 2/3\sqrt{3}$, and therefore
$\beta_c(n)\sim 3\sqrt{3}/4n$ as $n\to\infty$. The asymptotic vanishing of the critical value in the large $n$ limit suggests that most ABP will tend to stay close to the boundary of the well for any non-zero value of $\beta$, for hard-wall confinement, in agreement with existing observations~\cite{li2009accumulation,elgeti2015run,elgeti2009self,elgeti2013wall,vincent2025curvature}. 

\subsection{Shape transition in the positional distribution}
To understand the appearance of maxima and minima away from the origin (for $n>2$), in the vicinity of $\beta_c(n)$, define $\Delta \beta=\beta-\beta_c(n)$. Define $\Delta\xi_*=\xi_*-\Delta_n$ as the positions of the extrema, relative to $\Delta_n$. After substituting the expressions in Eq.~\ref{eq28} and expanding both l.h.s and r.h.s to the first non-vanishing order, we find that $\Delta\xi_*^2\propto \Delta\beta$, or 

\begin{equation}
\xi_*^{\pm}-\Delta_n\sim  ~\pm \alpha\sqrt{\Delta\beta}, 
\label{eq33}
\end{equation}
where the proportionality constant is 
\begin{equation}
\alpha=[(n-1)(n-2)(2n-1)\,\beta^2_c(n)]^{-\frac{1}{2}}\, \Bigg[\frac{n-2}{3n-4}\Bigg]^{-\frac{n-4}{4(n-1)}}. 
\label{eq:proportionality_const}
\end{equation}
and $\xi_*^{\pm}$ are the locations of the nontrivial extrema, with the $+$ sign representing the maximum and the $-$ sign, the minimum. The non-trivial extrema are thus predicted to be symmetrically located with respect to $\Delta_n$, in the limit of small $\Delta\beta$. 

Let us now show how the shape transitions are different for $n>2$ from $n=2$. For this, we compare transitions between $n=2$ and $n=4$ (as a specific example), shown in Fig.~\ref{fig:compare_peak_max_min_n_2_n_4}. In Fig.~\ref{fig:compare_peak_max_min_n_2_n_4}, the maxima, minima, and higher-order stationary points (HOSPs) are represented by star (red), circle (blue), and diamond-shaped markers, respectively, while the peak refers to the highest (or the tallest) point of the distribution, is indicated by a red $P$ symbol. If the HOSP at the origin is the peak, then the diamond symbol is colored red. 

In Fig.~\ref{fig:compare_peak_max_min_n_2_n_4}\textbf{a}, for $n=2$, all star-shaped points (red) are both maxima and peak simultaneously. For low values $\beta$, $\xi_*=0$ is a maximum (the distribution has a concave shape at this point~\cite{nakul2023stationary}). As $\beta$ increases and for $\beta \geq \beta_c$, another maximum appears at $\xi_*=({1-\frac{1}{3\beta}})^{1/2}$, which is zero at $\beta=\beta_c$. For $\beta>\beta_c$, the previous maximum (trivial) becomes a minimum (the concave shape becomes convex) and remains a minimum thereafter. Meanwhile, the new maximum (non-trivial) continuously shifts toward $\xi=1$ as $\beta$ increases. 

For $n=4$ (see Fig.~\ref{fig:compare_peak_max_min_n_2_n_4}\textbf{b}), the trivial extremum $\xi_* =0$ is always a HOSP, for any $\beta$. At low values of $\beta$, this extremum is also the peak, as indicated by a red diamond. As $\beta$ increases, $\Psi_n(\xi)$ becomes flatter in the vicinity of $\xi=0$ and when $\beta$ slightly exceeds a critical value $\beta_c(n)$, a pair of extrema appears spontaneously in the vicinity of $\xi=\Delta_4$ (Eq.~\ref{eq31}). Therefore, for $n\geq 4$, at $\beta$ slightly exceeding $\beta_c(n)$, $\xi_*$ has a non-zero value, unlike the case $n=2$. However, note that the newly generated maximum is not necessarily the highest point for $\Psi_n$ when $\Delta\beta$ is very small. However, as $\beta$ increases further, the new maximum eventually becomes the highest point, or peak and moves closer to $\xi=1$, while the minimum shifts toward $\xi=0$ at a much slower rate. Hereafter, we refer to the positions of the non-trivial maximum and minimum, i.e. $\xi_*^{\pm}$ (Eq.~\ref{eq33}) as the {\it shape parameters} of the distribution, with the added convention that $\xi_*^{\pm}\equiv 0$ for $\beta\leq \beta_c(n)$. Clearly, for $n=2$, $\xi_*^{-}=0$ for all $\beta$, while $\xi_*^{+}$ is predicted to grow continuously for $\beta>\beta_c(2)$. For $n\geq 4$, both shape parameters are predicted to undergo a discontinuous jump at 
$\beta=\beta_c(n)$.

\begin{table}[ht]
\caption{Comparison of theoretical predictions and simulation results of the gap parameter $\Delta_n$ and the critical activity $\beta_c(n)$ for ABP in power-law potential well.}
\label{Table-1}
\begin{center}
\begin{tabular}{ |c|c|c|c|c|c| } 
\hline
 & \rule{0pt}{2.5ex} $n$ & 2 & 4 & 8 & 20 \\
\hline
\multirow{2}{5em}{Theory} & \rule{0pt}{3ex} $\Delta_n$ & 0.0 & 0.7937 & 0.9175 & 0.9705 \\ 
\cline{2-6}
& \rule{0pt}{3ex} $\beta_c(n)$ & 1/3 & 0.3023 & 0.1595 & 0.0647 \\ 
\hline
\multirow{2}{5em}{Simulation} & \rule{0pt}{3ex} $\Delta_n$ & 0.0 & 0.7499 & 0.8920 & 0.9650 \\ 
\cline{2-6}
& \rule{0pt}{3ex} $\beta_c(n)$ & 0.3165 & 0.2942 & 0.1596 & 0.0686 \\ 
\hline
\end{tabular}
\end{center}
\end{table}
From the $n$-dependence of $\beta_c$, it follows that the shape transition in the positional distribution can also be observed when $n$ is varied at fixed $\beta$. We illustrate this in Fig.~\ref{fig:xi_max_vs_n_ABP_for_constant_beta_0_3}\textbf{a}, where $\beta=0.3$ is kept fixed, and $n$ is varied from 2 to 8, indicated by different colors/shapes. As $n$ increases, the peak at $\xi=0$ moves to a position close to $\xi=1$. For $n=2$, the peak at $\xi=0$ is a maximum (red star). As $n$ increases, for $n=4$, the peak remains in the same position but becomes a higher-order stationary point (HOSP, red diamond). When $n$ is increased further ($n=6$ and for $n=8$), the peak shifts discontinuously toward $\xi=1$. 

Fig.~\ref{fig:xi_max_vs_n_ABP_for_constant_beta_0_3}\textbf{b} illustrates the evolution of the locations of the extrema with change in  $\beta$. The intersection of $w_n(\xi_*)$ and $G(\beta,n)$ occurs at the nontrivial extrema. For a given $n$, if $w_n$ (solid lines) does not intersect with the constant line $G(\beta,n)$ (dashed lines), then the extrema is $\xi_* = 0$, which is either a maxima (red star) or a HOSP (red diamond). For $n=2$ and 4, the peak of the distribution is at $\xi=0$, which is a maximum for $n=2$ and a HOSP for $n=4$. As $n$ increases, $w_n(\xi)$ and $G(\beta,n)$ intersect, resulting in the appearance of a pair of extrema, as discussed above. Subsequently, the peak shifts either continuously (for $n=2$) or discontinuously (for $n\geq 4$) from $\xi=0$ to the newly formed maximum (red star) $\xi_*^{+}$.

In the next subsection, we subject our predictions to detailed quantitative verification in numerical simulations. This is necessary in view of the Gaussian assumption for the orientation variable $\chi$, which we have used in the simplification of the Fokker-Planck equation. 

\begin{figure*}[]
\centering
  \includegraphics[scale=0.54]{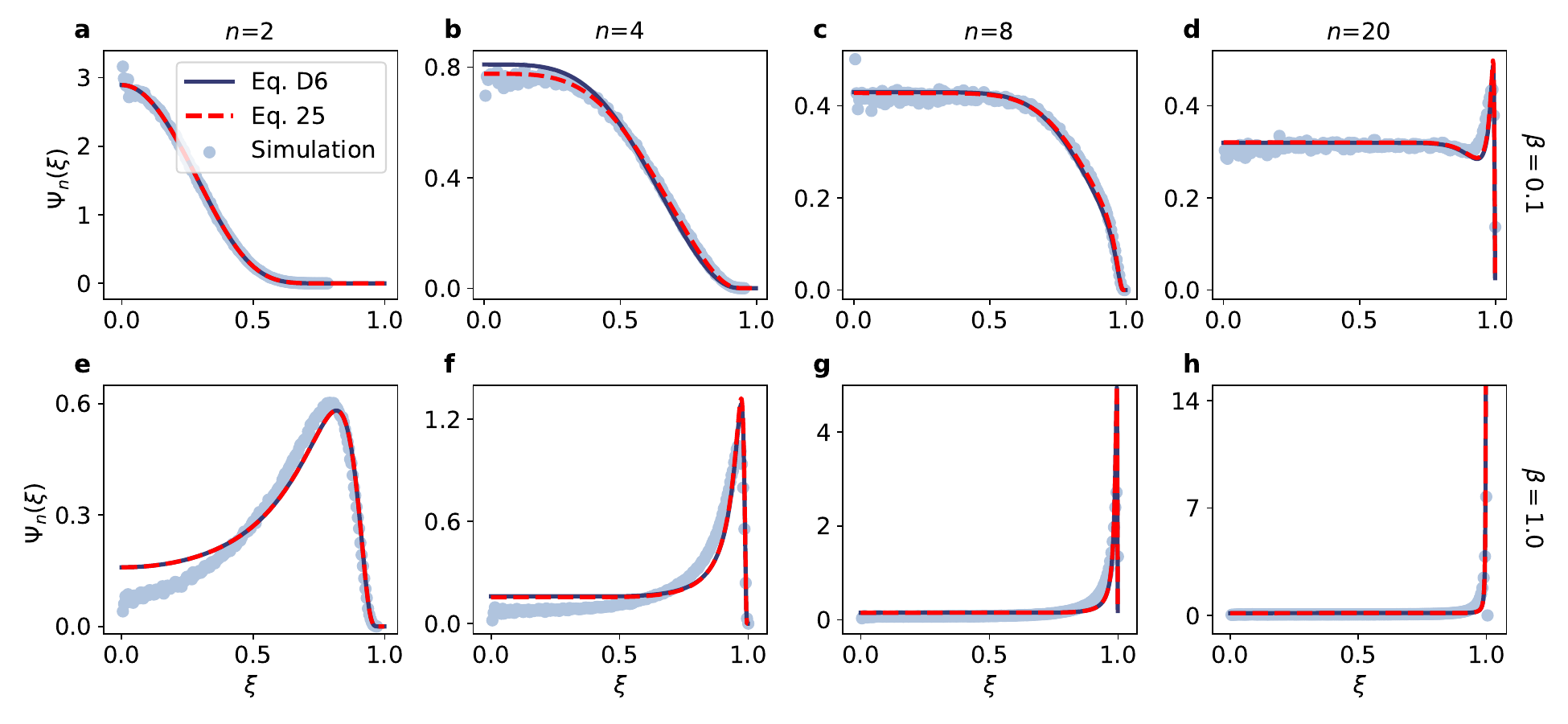}
  \caption{In both panels, $\Psi_n(\xi)$ obtained in simulations (gray circles) is plotted against $\xi$ for two different $\beta$ ($\beta=0.1$ in the top panel and $\beta=1.0$ in the bottom panel) and four different values of $n$ (2,4,8 and 20). As $n$ is increased, there is a shift in the position of the peak of the distribution function $\Psi_n(\xi)$ from the origin towards $\xi=1$ as predicted. The dashed lines (red) are theoretical fits using the expression in Eq.~\ref{expression_of_psi} and solid lines (dark blue) are approximate closed form expression of $\Psi_n$ for large $n$ (Eq.~\ref{closed_form_expression_large_n}). Note that both the theoretical predictions closely align with the numerical results.
  }
    \label{fig:Psi_at_diif_beta_for_diff_n}
\end{figure*}

\begin{figure*}[]
\centering
  \includegraphics[scale=0.45]{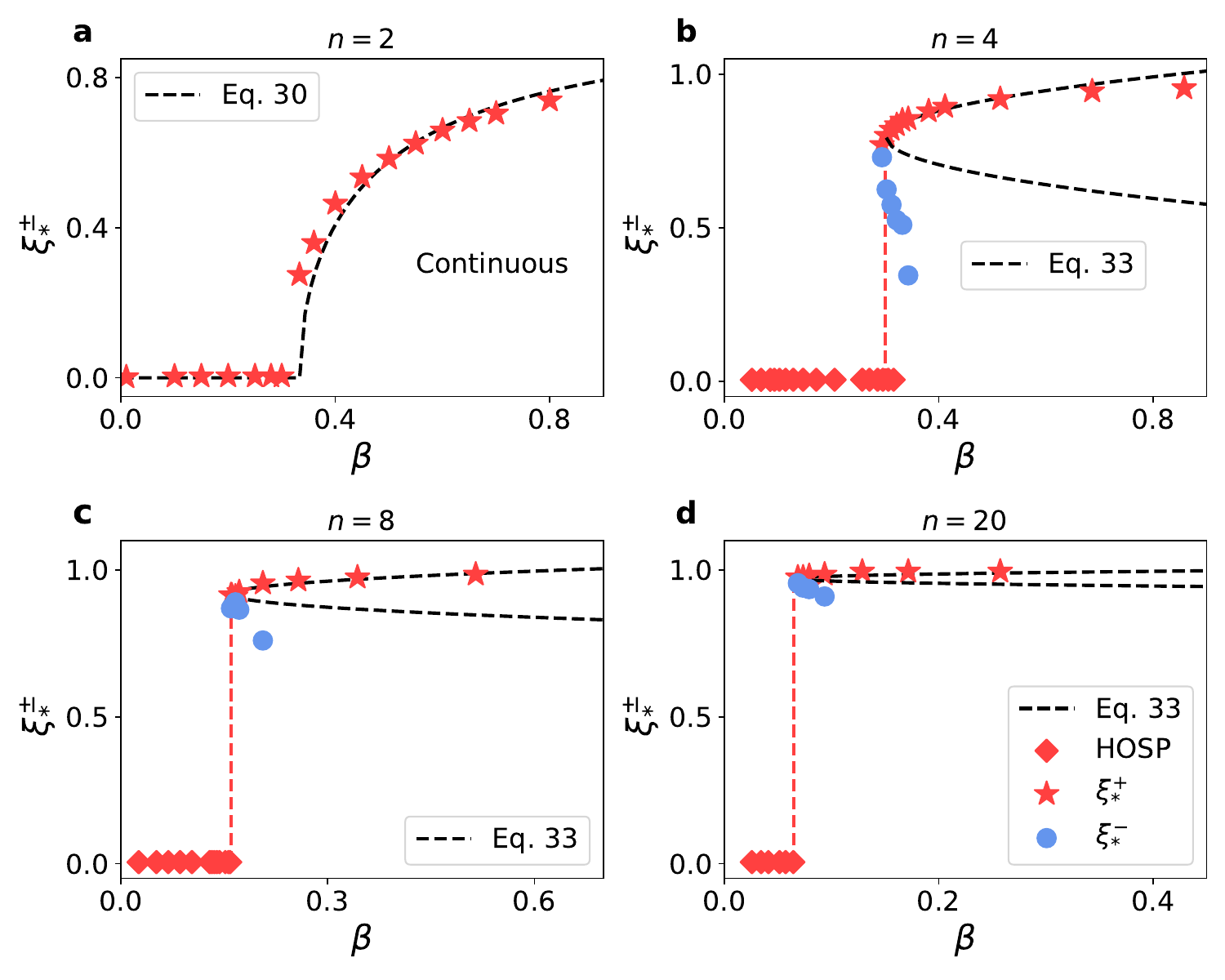}
  \caption{The figure demonstrates the shape transition in $\Psi_n(\xi)$ as a function of $\beta$ (ABP). The positions of the extrema $\xi_{*}^{\pm}$ of $\Psi_n(\xi)$, as obtained from the numerical simulations are plotted against $\beta$.  The black dotted lines are theoretical predictions (Eq.~\ref{eq30} for $n=2$ and Eq.~\ref{eq33} for $n>2$), detailed in the text), while the red dotted lines highlight the discontinuous jumps predicted by the theory (Eq.~\ref{eq31}) for $n>2$. The shape transition is continuous for $n=2$ and discontinuous for $n>2$ (see text).
  }
  \label{fig:xi_max_vs_beta_ABP_for_multiple_n}
\end{figure*}

\subsection{Numerical simulation results} 
We solve the overdamped Langevin equation (Eq.~\ref{eq:overdamped_langvin_eq}) numerically by the Euler-Maruyama method with $D_t=0$. Eq.~\ref{eq:overdamped_langvin_eq} is discretised as 
\begin{equation}\label{eq:discritised_ABP_eq}
\begin{aligned}
    x_{j+1} = x_j + u_0\cos\theta_j\, \Delta t - u_0^n\,\Bigl(\frac{\gamma_t}{\epsilon_0}\Bigl)^{n-1}\, r^{n-2}_j \, x_j \, \Delta t  \\[8pt]
    y_{j+1} = y_j + u_0\sin\theta_j\, \Delta t - u_0^n\,\Bigl(\frac{\gamma_t}{\epsilon_0}\Bigl)^{n-1}\, r^{n-2}_j \, y_j \, \Delta t \\[8pt]
    \theta_{j+1} = \theta_j + \sqrt{2 D_r\, \Delta t}\,\, W_{\theta , j},
\end{aligned}
\end{equation}
where $r_j=\sqrt{x_j^2 + y_j^2}$ and $W_{\theta,j}$ is a normal distribution with unit variance. In all simulations, $\Delta t=10^{-3}$ s, and averaged over $10^6$ particles. The parameters that are kept fixed in the simulations are shown in Table~\ref{Table-2}. Typically, the dimension of a Janus particle is of the order of 1 $\mu m$ and the maximum speed is $\sim 3\, \mu m/sec$~\cite{bechinger2016active}. For such a particle, at room temperature $T=300\,K$, $D_r\simeq$ 1.308, $s^{-1}$ and $\gamma_t=10^{-5}$ pNs nm$^{-1}$. 
\begin{table}[ht]
\begin{center}
\caption{The fixed simulation parameters for ABP in a power-law potential well. For explanations of the values, see text.}
\label{Table-2}
\renewcommand{\arraystretch}{1.6}  
\setlength{\tabcolsep}{6pt}  
\vspace{6pt}
\begin{tabular}{ |c|c|c| } 
\hline
 $u_0$ & $\gamma_{t}$ & $D_r$ \\
 nm s$^{-1}$ & pN s nm$^{-1}$ & rad$^2$ s$^{-1}$ \\
\hline
3670 & $ 10^{-5}$ & 1.308 \\ 
\hline
\end{tabular}
\end{center}
\end{table}
In all the simulations, we change $\beta$ by varying $\epsilon_0$. The other parameters are held fixed. 
\subsubsection{Positional distribution and shape parameters}
In Fig.~\ref{fig:Psi_at_diif_beta_for_diff_n}, the numerically obtained positional probability distribution $\Psi_n(\xi)$ is shown for different values of $n$ and $\beta$, along with the corresponding theoretical predictions. In both panels, the gray circles indicate the simulation data, obtained by solving Eq.~\ref{eq:discritised_ABP_eq}, while the dotted red line and solid dark blue line are the theoretical  expressions from Eq.~\ref{expression_of_psi} and Eq.~\ref{closed_form_expression_large_n}, respectively. Although Eq.~\ref{closed_form_expression_large_n} provides a closed form approximation of $\Psi_n(\xi)$ at large $n$, it still coincides almost perfectly with the more general expression for the same in Eq.~\ref{expression_of_psi}. It is important to note that the simulation data aligns well with the theoretical fit curves, apart from some small portions of the distribution near $\xi=0$, for large $\beta$ and small $n$. This suggests that even though the angular variable $\chi$ was treated as a Gaussian variable in the theoretical calculations, the predictions closely align with the simulation results. The simulations also confirm the existence of closely spaced maximum and minimum near $\xi=1$, clearly visible in Fig.~\ref{fig:Psi_at_diif_beta_for_diff_n}\textbf{d}, and the trends agree with the theoretical predictions. 

Next, we look at the shape transitions more closely. Fig.~\ref{fig:xi_max_vs_beta_ABP_for_multiple_n} \textbf{a}, \textbf{b}, \textbf{c} and \textbf{d} show how the shape parameters change as functions of $\beta$, for different values of $n$ (2, 4, 8 and 20 respectively).  In all figures, red stars and blue circles are the maxima and the minima of the distribution $\Psi_n(\xi)$ respectively, while the diamond-shaped markers are the higher order stationary points (HOSPs), colored red if they are peaks. 

\begin{figure*}[]
\centering
  \includegraphics[scale=0.56]{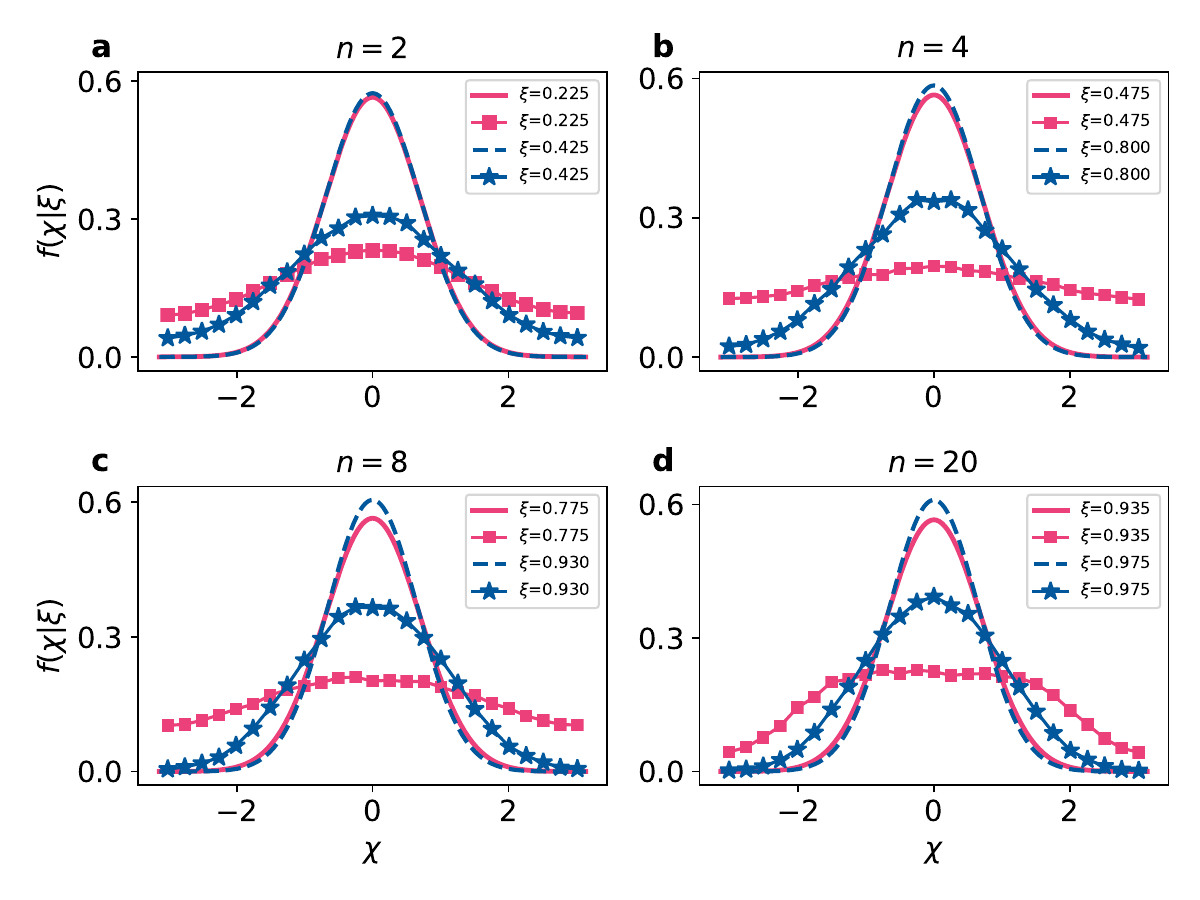}
  \caption{The conditional angular probability distribution $f(\chi|\xi)$ is plotted against $\chi$ for $\beta=0.1$ and $n=2,4,8$ and $20$. The theoretical fits (solid and dashed lines) are Gaussian distributions with variance given by Eq.~\ref{eq23}.}
  \label{fchi_distribution_numrical_beta_0_1}
\end{figure*}

\begin{figure*}[]
\centering
  \includegraphics[scale=0.56]{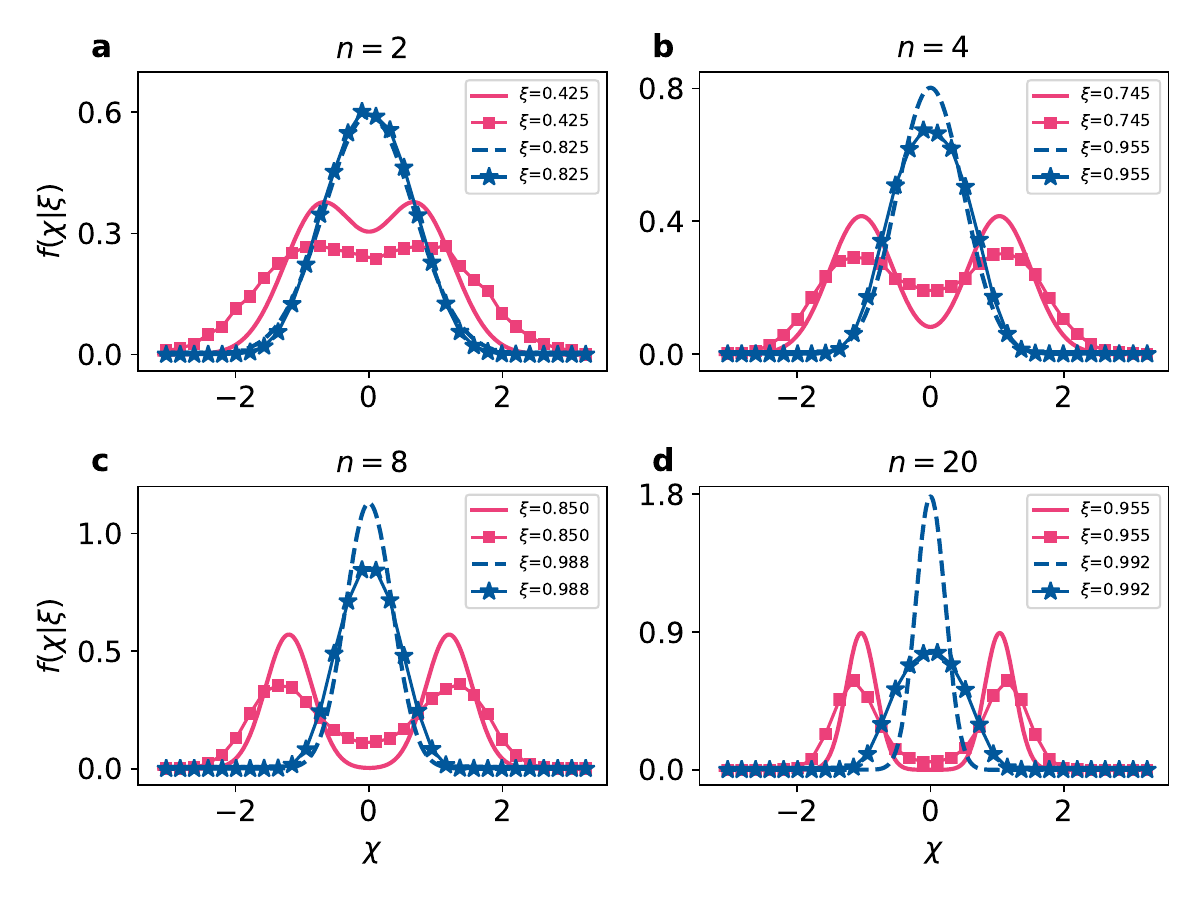}
  \caption{For $\beta=1.0$, the conditional angular probability distribution $f(\chi|\xi)$ is plotted against $\chi$ for $n=2,4,8$ and $20$. The distribution is unimodal with peak at $\chi=0$ close to $\xi=1$, but becomes bimodal with two symmetrically located peaks away from it. The theoretical fits are given by the expression in Eq.~\ref{eq:GAUSSIAN1}.}
  \label{fchi_distribution_numrical_beta_1_0}
\end{figure*}

\begin{figure}[]
\centering
  \includegraphics[scale=0.40]{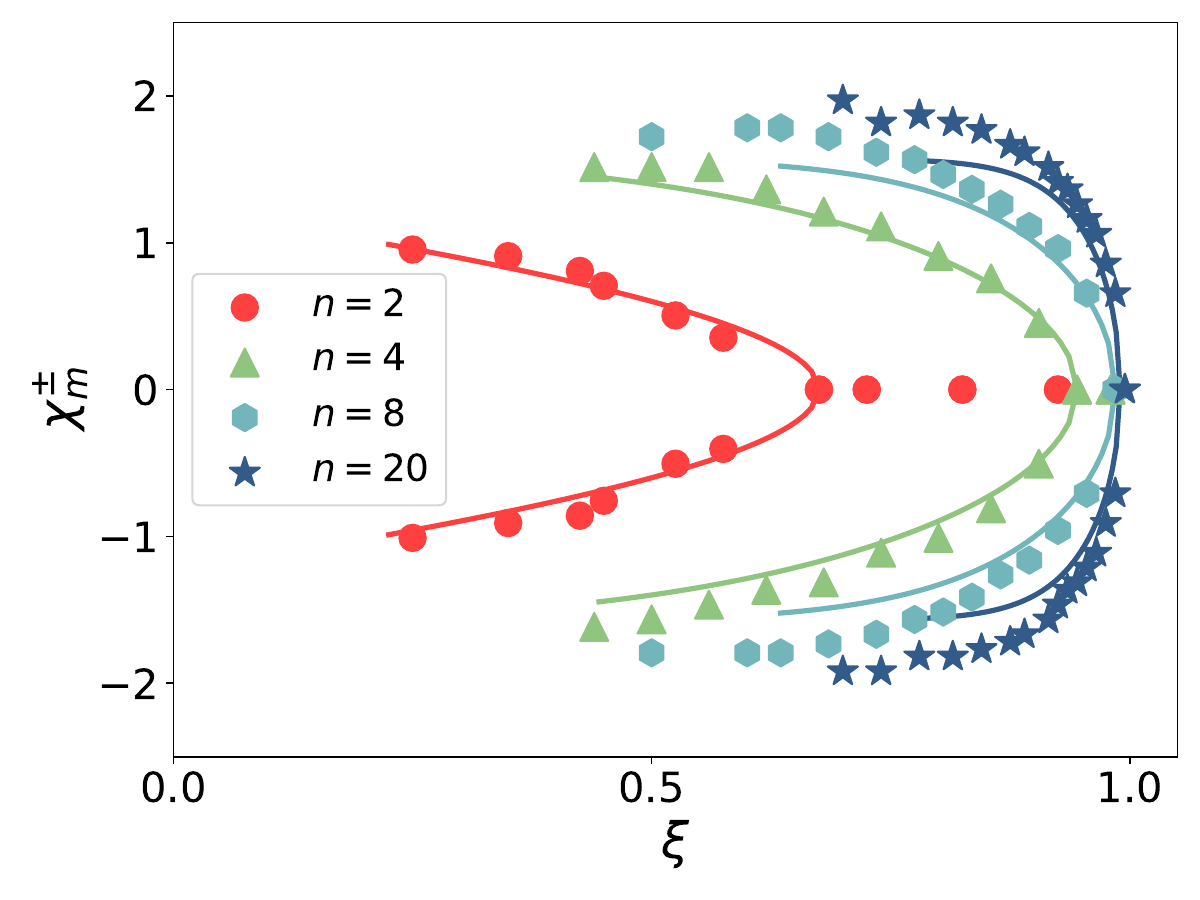}
  \caption{The bifurcation diagram for the angular distribution. For $\beta=1.0$, $\chi_{m}^{\pm}$ obtained from the simulation data (Fig.~\ref{fchi_distribution_numrical_beta_1_0}) are plotted as functions of $\xi$.  The solid lines (similar color for corresponding $n$) represent the theoretical prediction $\chi_{m}^{\pm}=\cos^{-1}{\tilde \xi}^{n-1}$, where ${\tilde \xi}$ is defined in Eq.~\ref{SHIFT}.}
  \label{fchi_distribution_max_vs_xi_beta_1_0}
\end{figure}
\subsubsection{Angular distribution}
Next, we shift our focus to the (conditional) angular probability distribution $f(\chi|\xi)$ (Eq.~\ref{eq9}). In Fig.~\ref{fchi_distribution_numrical_beta_0_1}, the distributions $f(\chi|\xi)$ are plotted against $\chi$ for four different values of $n$ (2,4,8 and 20) and the different symbols (or colors) correspond to different values of $\xi$. For all cases, $\beta=0.1$ and $D_t=0$. Regardless of the shape of the corresponding positional distributions for each $n$ (Fig.~\ref{fig:Psi_at_diif_beta_for_diff_n}\textbf{a},~\ref{fig:Psi_at_diif_beta_for_diff_n}\textbf{b},~\ref{fig:Psi_at_diif_beta_for_diff_n}\textbf{c} and ~\ref{fig:Psi_at_diif_beta_for_diff_n}\textbf{d}), all the sub-figures in Fig.~\ref{fchi_distribution_numrical_beta_0_1} show Gaussian-like unimodal distributions with peak at $\chi=0$, corresponding to radially outward orientation. As $\xi$ is reduced and we move towards the center, the peak at $\chi=0$ is suppressed and $f(\chi|\xi)$ becomes flatter, indicating a shift towards uniform orientation of the particles. With a change in $n$, for similar $\xi$, the peak of $\Psi_n(\xi)$ is broader for larger $n$ and the range of $\xi$ for finite $\Psi_n(\xi)$ increases toward $\xi=1$. As a result, the angular distributions cover a larger range of values of $\xi$, while keeping the qualitative shapes of $f(\chi|\xi)$ unchanged. For large value of $n$ (eg., $n=20$), although the peak of $\Psi_{20}(\xi)$ is close to $\xi=1$, the corresponding $f(\chi|\xi)$ remains almost unimodal (Fig.~\ref{fchi_distribution_numrical_beta_0_1}\textbf{d}). If we revisit the positional distribution for $n=20$ (Fig.~\ref{fig:Psi_at_diif_beta_for_diff_n}\textbf{d}), we find that $\Psi_{20}(\xi)$ is flat for most of $\xi$. Interestingly, for the same range of $\xi$, $f(\chi|\xi)$ is also nearly flat, indicating uniform orientation of ABP.

The conditional angular distribution for a higher value of $\beta$ ($\beta=1.0$) is shown in Fig.~\ref{fchi_distribution_numrical_beta_1_0}. The distinct difference between these and the previous  plots(Fig.~\ref{fchi_distribution_numrical_beta_0_1}) is that, for most values of $\xi <1$ the angular distribution is bimodal. Recall that the corresponding positional distributions peak near the trap radius $\xi=1$ (see Fig.~\ref{fig:Psi_at_diif_beta_for_diff_n}) and vanish rapidly away from the same. Hence, the angular distributions are plotted for $\xi$ close to 1. For small $n$, when a particle is located close to $\xi=1$, the angular distribution remains unimodal and the most probable orientation is radially outward. However, a little away from $\xi=1$, the distribution becomes bimodal. The symmetrically located peaks, denoted by $\chi_{\rm m}^{\pm}$, of the bimodal distribution, gradually move outward as $\xi$ is reduced. The value of $\xi$, where the bifurcation begins is defined as the {\it bifurcation point}, denoted by $\xi_c$. For $n=2$, we find that $\xi_c \approx 0.675$. For higher values of $n$, this bifurcation point moves closer to $\xi=1$ (see TABLE.~\ref{Table-3}). 

\begin{table}[ht]
\caption{Numerical values of bifurcation point $\xi_c$ for the unimodal-bimodal transition in the angular distribution, at $\beta=1.0$.}
\label{Table-3}
\begin{center}
\begin{tabular}{ |c|c|c|c|c| } 
\hline
 \rule{0pt}{3ex} $n$ & 2 & 4 & 8 & 20 \\
\hline
\rule{0pt}{3ex} $\xi_c$ & 0.675 & 0.945 & 0.985 & 0.990 \\ 
\hline
\end{tabular}
\end{center}
\end{table}

In Fig.~\ref{fchi_distribution_max_vs_xi_beta_1_0} we plot the angles $\chi_{\rm m}^{\pm}$ corresponding to the peaks of the bimodal angular distribution, plotted against $\xi$, with the convention that $\chi_{\rm m}^{\pm}=0$ for the unimodal distribution (with a peak at the origin). Simulation values for $\chi_{\rm m}^{\pm}$ are indicated by different symbols (or colors) for different $n$. The simulation data shows that as we move inward from $\xi=1$, the unimodal-bimodal transition in $f(\chi|\xi)$ starts at a specific position $\xi=\xi_c(n,\beta)$ (see Table~\ref{Table-3}). 

\subsection{Bifurcation theory for the angular distribution}

After incorporating the radial symmetry and by making use of Eq.~\ref{eq:B1} and Eq.~\ref{eq:B2}, the complete stationary state Fokker Planck equation in Eq.~\ref{eq:SFPE} can be expressed in the form 
\begin{equation}\label{SFPE_for_angular_dis_chi}
    \begin{split}
\frac{\partial}{\partial \xi}(v_{r}{\mathcal P})=\frac{1}{\beta}\frac{\partial^2{\mathcal P}}{\partial \chi^2}+\frac{1}{\xi}\bigg\{\frac{\partial}{\partial \chi}(v_{\phi}{\mathcal P})-v_r{\mathcal P}\bigg\}
\end{split}
\end{equation}
where $v_r=V_r/u_0, v_{\phi}=V_{\phi}/u_0$ are polar components of the dimensionless velocity ${\bf v}={\bf V}/u_0$, and are given by the explicit expressions
\begin{equation}
v_r=\cos\chi-\xi^{n-1}~~~;~~~v_{\phi}=\sin\chi.
\label{eq37}
\end{equation}
After substituting Eq.~\ref{eq37} in Eq.~\ref{SFPE_for_angular_dis_chi} and performing some simplifications, we arrive at the equation
\begin{equation}
(\cos\chi-\xi^{n-1})\frac{\partial {\mathcal P}}{\partial \xi}=\frac{1}{\beta}\frac{\partial^2{\mathcal P}}{\partial \chi^2}+\frac{\sin\chi}{\xi}\frac{\partial {\mathcal P}}{\partial \chi}+n\xi^{n-2}{\mathcal P}.
\label{eq:NFPE}
\end{equation}

The emergence of the bimodal angular distribution when $\beta>\beta_c$, as observed in the simulations (Fig.~\ref{fchi_distribution_numrical_beta_1_0}) may be understood using the following argument. Define the angle $\chi^{*}(\xi)$ such that the radial velocity in Eq.~\ref{eq37} is expressed as 
\begin{equation}\label{eq:angular_Theory3}
\begin{aligned}
    v_r(\xi,\chi)= \cos\chi-\cos\chi^*(\xi)~~;~~\chi^*(\xi)=\cos^{-1}\xi^{n-1}.
\end{aligned}
\end{equation}
From the previous equation, it follows that $v_r(\xi,\chi^*(\xi))=0$. Next, we define the angular deviation 
$\chi^{\prime}=\chi\pm\chi^*(\xi)$ where $\chi^*(\xi)=\cos^{-1}(\xi^{n-1})$ is taken to be positive. Consider the probability distribution ${\mathcal P}(\xi,\chi)\to \Psi(\xi,\chi^{\prime})$ in Eq.~\ref{eq:NFPE}. Then the derivatives in that equation transform as 
\begin{equation}\label{transformation_eqs}
    \begin{split}
      & \frac{\partial}{\partial \xi}\to \frac{\partial}{\partial \xi} - \frac{\partial \chi^*}{\partial \xi}\,\frac{\partial}{\partial \chi^{\prime}},\\
      & \frac{\partial}{\partial \chi}\to \frac{\partial}{\partial \chi^{\prime}}. 
    \end{split}
\end{equation}
Also, for small $\chi^{\prime}$, the radial and tangential velocity components in Eq.~\ref{eq37} may be approximated as 
\begin{equation}\label{Vr_V_phi_for_small_chi_prime}
    \begin{split}
      & v_r\approx -\chi^{\prime}\, \sin\chi^*,\\
      & v_{\phi}\approx \sin\chi^* + \chi^{\prime}\,\cos\chi^*.
    \end{split}
\end{equation}

After substituting Eq.~\ref{Vr_V_phi_for_small_chi_prime} in Eq.~\ref{eq:NFPE} the SFPE becomes 
\begin{equation}\label{eqE5}
    \begin{split}
      & -\chi^{\prime} \sin\chi^* \, \frac{\partial \Psi}{\partial \xi} + \chi^{\prime} \sin\chi^* \, \frac{\partial \chi^*}{\partial \xi} \frac{\partial \Psi}{\partial \chi^{\prime}} - \frac{\sin\chi^*}{\xi}\,\frac{\partial \Psi}{\partial \chi^{\prime}}  \\
      & - \frac{\cos\chi^*}{\xi}\,\frac{\partial (\chi^{\prime} \Psi)}{\partial \chi^{\prime}} - \Psi\,\frac{\partial \chi^*}{\partial \xi} \big(\chi^{\prime} \cos\chi^* - \sin\chi^*\big) \\
      & - \frac{\chi^{\prime}\Psi}{\xi}\, \sin\chi^* = \frac{1}{\beta}\, \frac{\partial^2 \Psi}{\partial {\chi^{\prime}}^2}\,.
    \end{split}
\end{equation}

For $\xi$ close to 1, an explicit expression for $\chi^*(\xi)$ can be found as follows. Given that $\chi^*(1)=0$, for $1-\xi\ll 1$, we expect $\chi^*(\xi)$ to be small, and hence the approximation $\cos \chi^*\simeq 1-(\chi^*)^2/2$ can be used. Also, for $1-\xi\ll 1$, $1-\xi^{n-1}\simeq (n-1)(1-\xi)$ after using the appropriate binomial expansion. Upon using these limiting expressions in Eq.~\ref{eq:angular_Theory3}, we find 
\begin{equation}
\chi^*(\xi)\sim c(1-\xi)^{\alpha}~~~~~(1-\xi)\ll 1
\label{eq:ALPHA}
\end{equation}
where, our predictions are $c=\sqrt{2(n-1)}$ and $\alpha=1/2$. 
 
Hence, in the limit $\xi\to 1$, we may approximate $\sin\chi^*\simeq \chi^*$ and $\cos\chi^*\simeq 1$ in Eq.~\ref{eqE5}, which leads to the simplified equation

\begin{equation}
-\chi^{\prime}\frac{\partial}{\partial\xi}(\chi^*\Psi)
=\frac{1}{\beta}\frac{\partial^2\Psi}{\partial\chi^{{\prime}^2}}+\gamma(\xi)\frac{\partial}{\partial \chi^{\prime}}(\chi^{\prime}\Psi)
\label{eq:XFPE}
\end{equation}
where 
\begin{equation}
\gamma(\xi)=\xi^{-1}-\frac{1}{2}\frac{d}{d\xi}(\chi^*)^2, 
\label{eq:GAMMA}
\end{equation}
for which the theoretical prediction (corresponding to the theoretical values for $c$ and $\alpha$) is 
\begin{equation}
\gamma_{\rm th}(\xi)=n-1+\xi^{-1},
\label{eq:GAMMA_THEORY}
\end{equation}
and $\gamma_{\rm th}\to n$ as $\xi\to 1$. 
The more general expression is 
\begin{equation}
\gamma(\xi)=\xi^{-1}+\alpha c^2(1-\xi)^{2\alpha-1}, 
\label{eq:GAMMA-SIM}
\end{equation}
and if $\alpha>1/2$, $\gamma\to 1$ as $\xi\to 1$ whereas if $\alpha<1/2$, $\gamma$ tends to diverge as $\xi\to 1$. 

In order to solve Eq.~\ref{eq:XFPE}, let us define $F(\chi^{\prime},\xi)=\chi^*\Psi$, which satisfies the equation
\begin{equation}
-\chi^*\chi^{\prime}\frac{\partial F}{\partial \xi}=\gamma(\xi)\frac{\partial}{\partial \chi^{\prime}}(\chi^{\prime}F)+ \frac{1}{\beta}\frac{\partial^2 F}{\partial{\chi^{\prime}}^2}
\label{eq:FR}
\end{equation}
The l.h.s of the above equation vanishes as $\xi\to 1$, hence in this limit, $F(\chi,\xi)$ is given by the expression 
\begin{equation}
F(\chi^{\prime},\xi)\propto \exp\bigg(-\frac{\beta\gamma(\xi)(\chi^{\prime})^2}{2}\bigg)
\label{eq:GAUSSIAN}
\end{equation}
The proportionality constant in Eq.~\ref{eq:GAUSSIAN} may be chosen such that the function $F(\chi^{\prime},\xi)$ is normalised, i.e., $\int F(\chi^{\prime},\xi)d\chi^{\prime}=1/2$ (the normalisation factor is 1/2 instead of 1 because of the degeneracy in the definition of $\chi^{\prime}$). For sufficiently small variance, the limits of integration may be taken as $-\infty$ and $\infty$. After normalization, the function in Eq.~\ref{eq:GAUSSIAN} gives the conditional distribution for the orientation angle $\chi$:
\begin{equation}
f(\chi|\xi)\simeq \frac{1}{2}\sqrt{\frac{\beta \gamma(\xi)}{2\pi}}\left\{e^{-\frac{\beta\gamma(\xi)}{2}(\chi-\chi^*)^2}+  e^{-\frac{\beta\gamma(\xi)}{2}(\chi+\chi^*)^2}\right\}
\label{eq:GAUSSIAN1}
\end{equation}
which is fitted to the numerical data in Fig.~\ref{fchi_distribution_numrical_beta_1_0} using the expression for $\gamma(\xi)$ in Eq.~\ref{eq:GAMMA_THEORY}.

The bifurcation theory thus predicts that $\chi_m^{\pm}=\pm \chi^*$, with 
$\chi^*$ given by Eq.~\ref{eq:angular_Theory3}, which grows continuously as $\xi$ is reduced below 1 (Eq.~\ref{eq:angular_Theory3}). However, simulation data shows that bifurcation does not begin until $\xi=\xi_c(n,\beta)$. Therefore, for the sake of fitting the theory with simulation data, we introduce a shifted position variable
\begin{equation}
{\tilde \xi}=1-[\xi_c(n,\beta)-\xi]
\label{SHIFT}
\end{equation}
and replace $\xi\to {\tilde \xi}$ in Eq.~\ref{eq:angular_Theory3} for the theoretical fit curves in Fig.~\ref{fchi_distribution_max_vs_xi_beta_1_0}. The modified fit functions match well with the simulation data. The reason for this $n$-dependent shift in the origin of the angular bifurcation is not clear to us at this time. 


\begin{figure*}[]
\centering
  \includegraphics[scale=0.50]{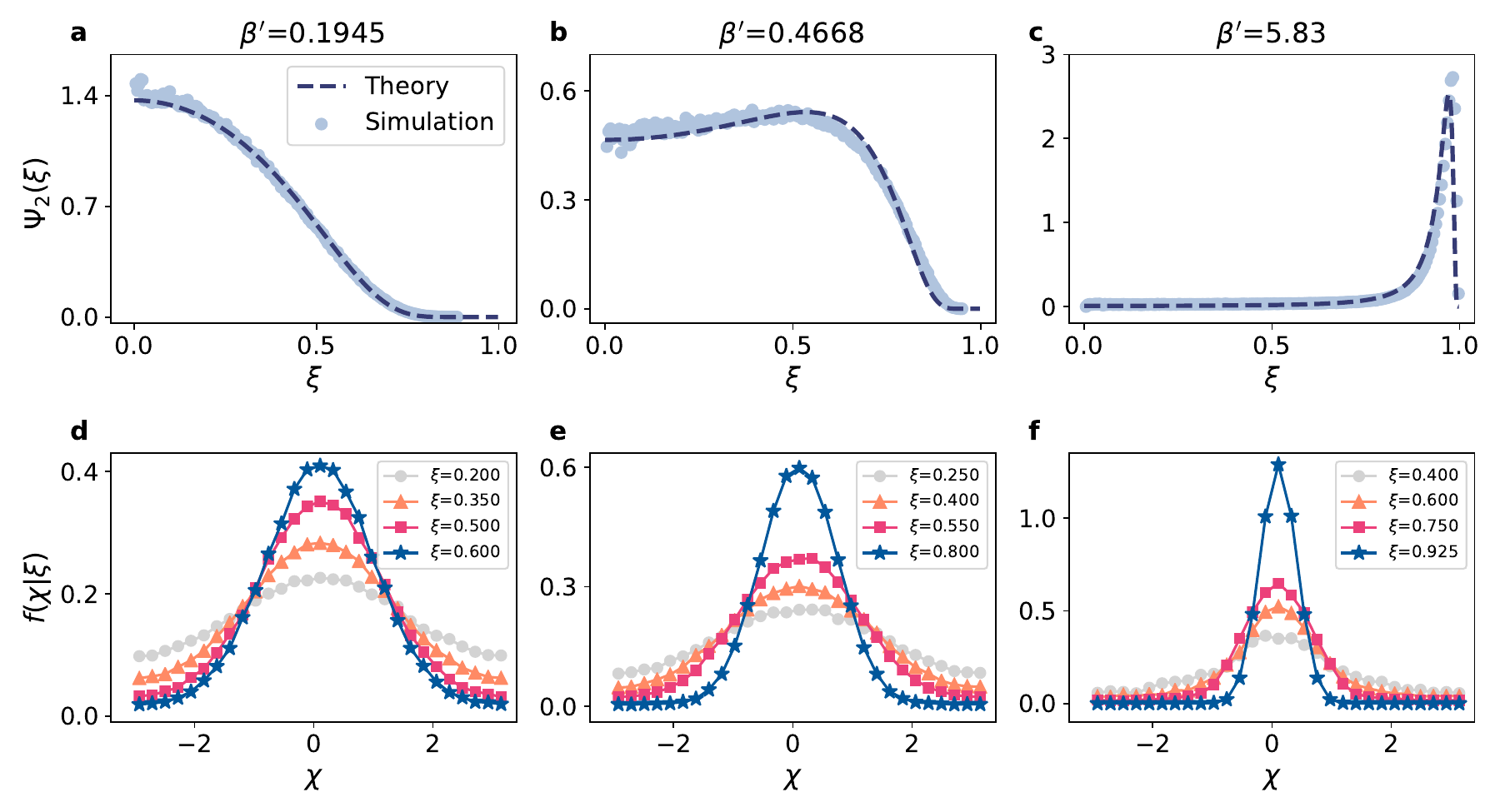}
  \caption{The upper and lower panels show the positional distributions $\Psi_2(\xi)$ and the angular distribution $f(\chi|\xi)$ respectively, of a RTP trapped inside a harmonic potential well. In the upper panel, the black dotted lines represent the theoretical prediction (Eq.~\ref{psi_harmonic_potential_RTP}) and the blue circles show the simulation data. Similar to ABP in a harmonic potential, a continuous shape transition is observed for the positional distribution, as predicted by theory. In the lower panel, the different colors (or symbols) represent the simulation data for the angular distributions at different $\xi$ for fixed $\beta^{\prime}$. Unlike ABP, the angular distribution remains nearly unimodal for RTP for all $\xi$.}
  \label{fig:Psi_chi_dist_RTP_dif_beta_for_n_2}
\end{figure*}

\begin{figure*}[]
\centering
  \includegraphics[scale=0.50]{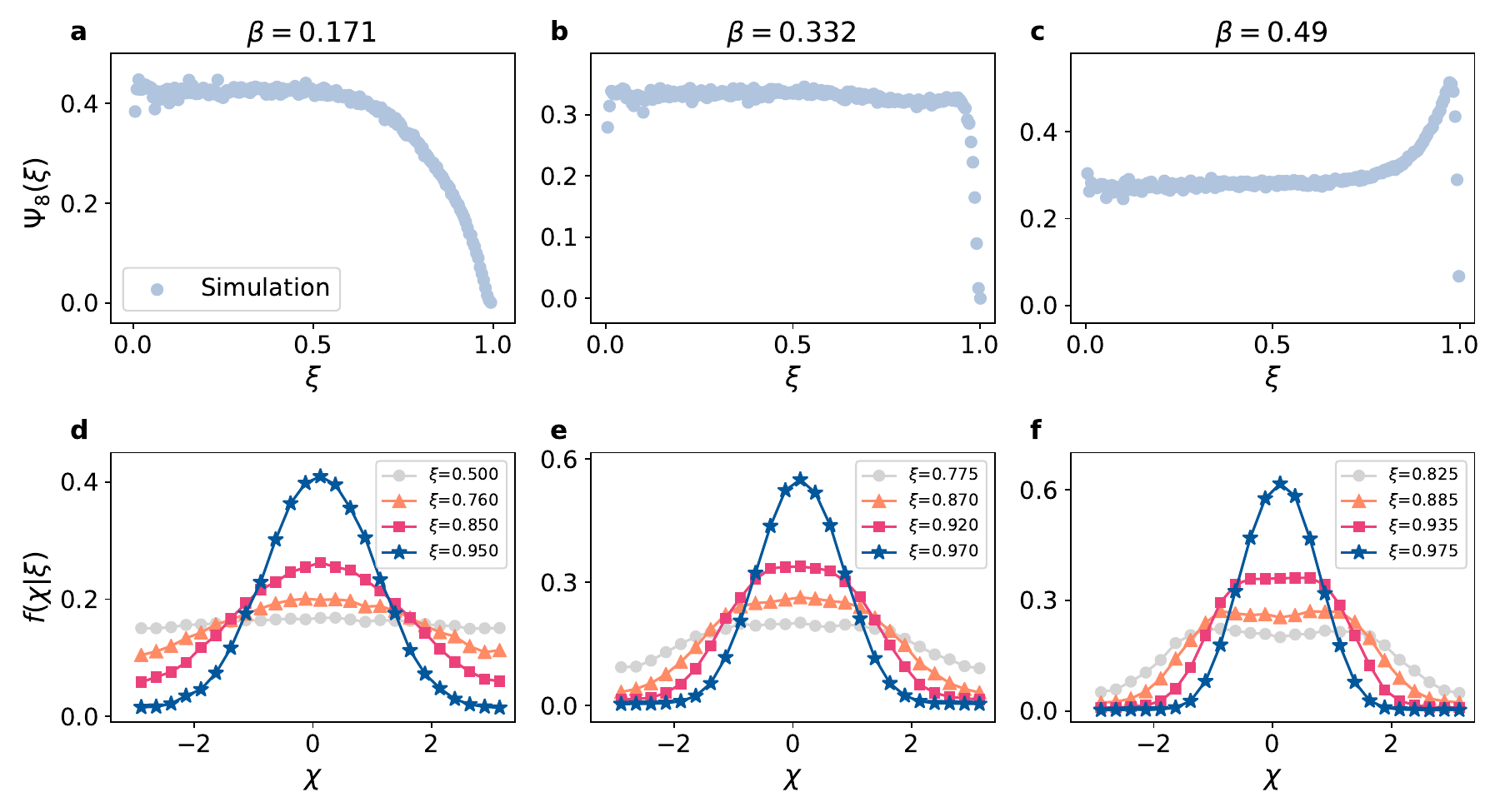}
  \caption{RTP in a power-law potential well with $n=8$. In the upper panel, the simulation data for the positional distribution $\Psi_8(\xi)$ is represented by blue circles and in the lower panel, the corresponding angular distributions $f(\chi|\xi)$ are shown for different values of $\xi$. Both sets of data are shown for three different values of $\beta$.}
  \label{fig:Psi_chi_dist_RTP_dif_beta_for_n_8}
\end{figure*}

\begin{figure*}[]
\centering
  \includegraphics[scale=0.42]{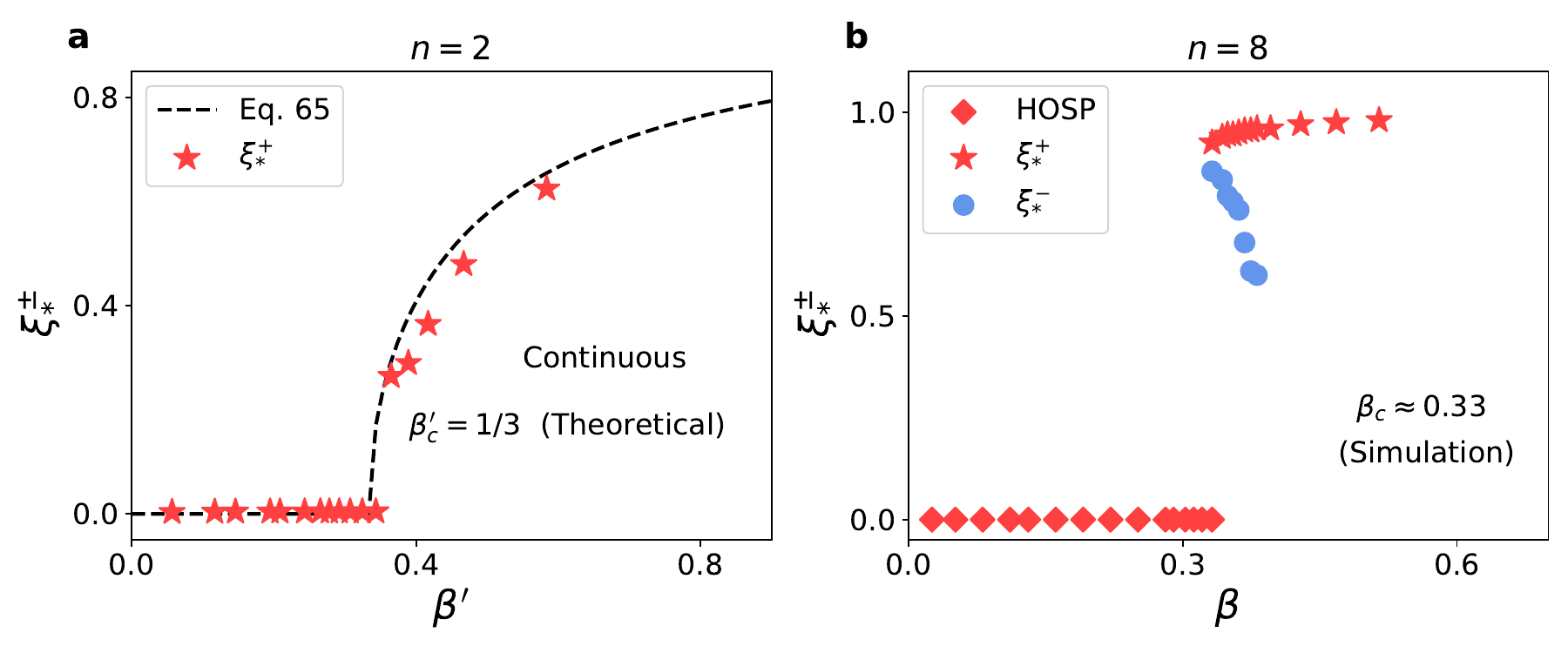}
  \caption{Shape parameter for RTP in a power-law potential well, for two different $n$. In part $\textbf{a}$, data for $n=2$ and in part $\textbf{b}$, data for $n=8$ are shown. All the symbols have similar meaning as in ABP (see Fig.~\ref{fig:xi_max_vs_beta_ABP_for_multiple_n}), The shape transition in $\textbf{a}$ is continuous and that $\textbf{b}$ is discontinuous.}
  \label{fig:xi_max_vs_beta_RTP_for_n_2_n8}
\end{figure*}

\section{Run and tumble particle in harmonic potential well}\label{sec:RTP}
In this section, we show how our formalism can be suitably modified to study the stationary states of a run-and-tumble particle (RTP) in a potential well. For the sake of brevity, we restrict the present discussion to the harmonic potential ($n=2$) only, for which the potential energy function is 
\begin{equation}
U(r)=\frac{\kappa}{2}r^2.
\label{harmonic_potential}
\end{equation}
The essential difference between a RTP and an ABP is that the former has, in additional to rotational and translational diffusion, an intrinsic tumbling mechanism which produces almost discontinuous changes in orientation over a very brief interval of time. In the absence of chemical gradients in the environment, the tumbles can be effectively described by a Poisson process characterised by a constant rate $\lambda$. 

\subsection{Fokker-Planck equation for RTP}

Let $W({\hat{\mathbf{u}}}|{\hat{\mathbf u}^{\prime}})$ denote the probability density for choosing the direction $\hat{\mathbf u}$ after a tumble event, given that the direction of motion before the tumble is $\hat{\mathbf u}^{\prime}$. The FPE which describes the motion of such a RTP follows from Eq.~\ref{FPE} by adding gain and loss terms describing the changes in orientation because of tumbles, and is given by 

\begin{equation}
    \begin{split}
            \frac{\partial P(\mathbf{r},\hat{\mathbf{u}},t)}{\partial t} = & - \boldsymbol{\nabla} \cdot \mathbf{J}_t + D_r {\nabla}^{2}_{\hat{\mathbf{u}}} P(\mathbf{r},\hat{\mathbf{u}},t) -\lambda P(\mathbf{r},\hat{\mathbf{u}},t) \\
            & + \lambda \int  W({\hat{\mathbf{u}}}|{\hat{\mathbf u}^{\prime}})P(\mathbf{r},\hat{\mathbf{u}}^{\prime},t)\, d\hat{\mathbf{u}}^{\prime},
    \label{FPE_RTP_harmonic_0}
    \end{split}
\end{equation}
where $\int W({\hat{\mathbf{u}}}|{\hat{\mathbf u}^{\prime}})d\hat{\mathbf{u}} =1$. We consider only the situation where the orientation angle after a tumble is uncorrelated with the same before it and chosen isotropically, so that $W({\hat{\mathbf{u}}}|{\hat{\mathbf u}^{\prime}}) = \Omega_d^{-1}$ where $\Omega_d$ is the solid angle in $d$ dimensions. After the substitution in the FPE, in stationary state, Eq.~\ref{FPE_RTP_harmonic_0} becomes  
\begin{equation}
            - \boldsymbol{\nabla} \cdot \mathbf{J}_t + D_r {\nabla}^{2}_{\hat{\mathbf{u}}} P(\mathbf{r},\hat{\mathbf{u}},t) -\lambda \bigg[P(\mathbf{r},\hat{\mathbf{u}},t)-\frac{\Phi (\mathbf{r})}{\Omega_d}\bigg]=0
    \label{FPE_RTP_harmonic}
\end{equation}
where the translational probability current density vector is ${\bf J}_t={\bf v}P$, with 
\begin{equation}
{\bf v}= u_0\hat{\mathbf{u}} -  \frac{\kappa}{\gamma_t } \, \mathbf{r}, 
\label{eq35}
\end{equation}
from which the natural length scale $R=\gamma_t u_0/\kappa$ is identified. As for the ABP, we define $\xi=r/R$ as the dimensionless distance.   

Integration of Eq.~\ref{FPE_RTP_harmonic} over ${\hat{\mathbf {u}}}$ leads to Eq.~\ref{non_dimensional_K_zero_eq} as in the case of ABP. This is because the gain and loss terms proportional to $\lambda$ cancel each other after this integration. Therefore, for $D_t=0$, we find 
\begin{equation}
\overline{\hat{\mathbf {u}}}\sim \xi~ \hat{\bf r}
\end{equation}
similar to ABP. To arrive at an equation for the positional probability distribution $\Phi(\mathbf{r})$, we integrate Eq.~\ref{FPE_RTP_harmonic} over $\hat{\mathbf{u}}$ after multiplying the equation by $\mathbf{v}$ (Eq.~\ref{eq35}), which leads to 
\begin{equation}\label{FPE_in_d_dimension_RTP}
    \begin{split}
   u_0^2 \int \hat{\mathbf{u}} (\hat{\mathbf{u}} \cdot \boldsymbol{\nabla} P) \, d^d \hat{\mathbf{u}} \, + S \, \frac{u_0^2}{R^2}\, \Phi\, \mathbf{r}\, - \frac{u_0^2}{R^2} \,\mathbf{r} (\mathbf{r} \cdot \boldsymbol{\nabla} \Phi)=0
    \end{split}
\end{equation}
where we have defined 
\begin{equation}
    S = \frac{\lambda}{\beta D_r} + (d-1) \, \frac{1}{\beta} - (d+1).
\end{equation}

In $d=2$, in terms of the scaled coordinate $\xi$, Eq.~\ref{FPE_in_d_dimension_RTP} is reduced to 
\begin{equation}\label{FP1_RTP}
\begin{split}
  &  \hat{\mathbf{r}} \cdot \int \hat{\mathbf{u}} \, (\hat{\mathbf{u}} \cdot \boldsymbol{\nabla}_{\xi} \mathcal{P}) \, d\theta = - \xi \big[ S\, \Psi -\, \xi \,\frac{\partial \Psi}{\partial \xi} \big] 
\end{split}
\end{equation}
which generalises Eq.~\ref{FP1} for RTP. The l.h.s in Eq.~\ref{FP1_RTP} is given by Eq.~\ref{FP2}, where we put $n=2$ and substitute in Eq.~\ref{FP1_RTP}, which takes us to the exact Fokker-Planck equation for the scaled radial coordinate of the RTP in harmonic potential:

\begin{equation}\label{FP3_RTP}
\begin{split}
  &  \sigma_{\cos \chi}^2  \frac{\partial \Psi}{\partial \xi}+\Bigl[\partial_{\xi} \sigma_{\cos \chi}^2 + \frac{2 \sigma_{\cos \chi}^2-1}{\xi}+ (4+S)\xi \Bigr]\Psi =0.
\end{split}
\end{equation}
After using the result in Eq.~\ref{eq23} which follows from the assumption of $\chi$ being Gaussian, and putting $n=2$, the following approximate Fokker-Planck equation for 
$\Psi$ is arrived at:
\begin{equation}\label{diffential_eq_FP_RTP}
       \frac{1}{2} \big(1 - \xi^{2}\big)^2 \, \frac{\partial \Psi}{\partial \xi} +\Bigl[S\, \xi + 3\, \xi^3 \Big]\Psi\simeq 0, 
\end{equation}
whose solution has the same mathematical form as that for the ABP given in Eq.~\ref{psi_harmonic_potential}, with the replacement $\beta\to \beta^{\prime}$, i.e., 
\begin{equation}\label{psi_harmonic_potential_RTP}
    \begin{split}
        & \Psi_2(\xi) = C_2\Bigl[1 -  \xi^{2} \Bigr]^{-3}\exp \Bigl[-\frac{ \xi^2}{\beta^{\prime} (1 -  \xi^2)} \Bigr], ~~~~({\rm RTP})
    \end{split}
\end{equation} 
where 
\begin{equation}
{\beta^{\prime}} = \beta \Big[1 + \frac{\lambda}{D_r}\Big]^{-1}.
\label{betaprime}
\end{equation}
Note that, if $\lambda=0$, $\beta^{\prime}=\beta$ as expected. The predicted shape transition for RTP is continuous, similar to that for ABP in a harmonic potential. This is because the former has a similar form of mathematical expression of $\Psi_2(\xi)$ as the latter, except that $\beta$ is modified to $\beta^{\prime}$. Hence, the critical value for the transition is given by the equation $\beta_c^{\prime}=1/3$, leading to 
\begin{equation}
\beta_c=\frac{1}{3}\Big[1 + \frac{\lambda}{D_r}\Big].~~~(n=2, {\rm RTP})
\label{BETACRTP}
\end{equation}
The positions of the maxima are given by 
\begin{equation}\label{continous_RTP}
\xi_*^{+}=\sqrt{1-1/3\beta^{\prime}}~~~~~~(\beta^{\prime}>\frac{1}{3})
\end{equation}
\subsection{Numerical simulation results}
We now discuss the numerical solutions for the RTP confined in a power-law potential well in $d=2$. The discretised Langevin equation for this system is a slightly modified form of Eq.~\ref{eq:discritised_ABP_eq}, with the tumble dynamics added~\cite{toschi2019flowing}: 

\begin{equation}\label{eq:discritised_RTP_eq}
\begin{aligned}
    x_{j+1} = x_j + Q_j\, u_0\cos\theta_j\, \Delta t - u_0^n\,\Bigl(\frac{\gamma_t}{\epsilon_0}\Bigl)^{n-1}\, r^{n-2}_j \, x_j \, \Delta t  \\[8pt]
    y_{j+1} = y_j + Q_j\, u_0\sin\theta_j\, \Delta t - u_0^n\,\Bigl(\frac{\gamma_t}{\epsilon_0}\Bigl)^{n-1}\, r^{n-2}_j \, y_j \, \Delta t \\[8pt]
    \theta_{j+1} = \theta_j + (1-Q_j)\,\Delta\theta_{\rm tumble} + \sqrt{2 D_r\, \Delta t}\,\, W_{\theta , j},
\end{aligned}
\end{equation}
where $Q_j$ is either 1 or 0, depending on whether the RTP is in run or tumble mode, respectively. $\Delta\theta_{\rm tumble}$ is the change in orientation angle as a result of a tumble event, which is randomly drawn from a uniform distribution in the range $[0, 2\pi]$. As tumbles are assumed to be a Poisson process with constant rate $\lambda$, the average number of tumbles that occur during a time interval $\Delta t$, is $\lambda \Delta t$. If $P_{\rm run}$ and $P_{\rm tumble}$ are the probabilities that a RTP only runs or only tumbles in a time interval $\Delta t$,  respectively, then $P_{\rm run}=e^{-\lambda \Delta t}$ and $P_{\rm tumble}=1-e^{-\lambda \Delta t}$ such that $P_{\rm run}+P_{\rm tumble}=1$. In every time step, the value of $Q_j$ is thus updated, which decides whether the RTP is in run or in tumble mode. The remaining terms in Eq.~\ref{eq:discritised_RTP_eq} are the same as in Eq.~\ref{eq:discritised_ABP_eq} for ABP, and the parameters have the same values as mentioned in Table~\ref{Table-2}. The tumble rate $\lambda$ is assigned the value 1 s$^{-1}$ for all simulations, and $\Delta t=10^{-3}$s. The total number of particles is $N=10^6$.

In Fig.~\ref{fig:Psi_chi_dist_RTP_dif_beta_for_n_2} (\textbf{a}, \textbf{b} and \textbf{c}), numerical results for the positional distribution $\Psi_2(\xi)$ is plotted against $\xi$ for three different values of $\beta^{\prime}$, where the black dotted lines indicate theoretical predictions (Eq.~\ref{psi_harmonic_potential_RTP}) and the blue circles are simulation data points. Similar to ABP, under the Gaussian assumption for the angle variable $\chi$, simulation data for $\Psi_2(\xi)$ agree well with the theory. As $\beta^{\prime}$ increases, the peak (or maximum) of the positional distribution continuously shifts from $\xi=0$ to $\xi=1$. Fig.~\ref{fig:Psi_chi_dist_RTP_dif_beta_for_n_2} (\textbf{d}, \textbf{e} and \textbf{f}), show the corresponding angular distribution $f(\chi|\xi)$ for $n=2$. As $\beta^{\prime}$ increases, peak of the distribution increases, however, it retains its Gaussian-like shape. Unlike  ABP, we do not see any bifurcation and bimodal peaks in the angular distribution of RTP, regardless of the value of $\beta^{\prime}$.

To compare the results with ABP, for $n>2$, numerical results for RTP are shown in Fig.~\ref{fig:Psi_chi_dist_RTP_dif_beta_for_n_8} where $n=8$. Fig.~\ref{fig:Psi_chi_dist_RTP_dif_beta_for_n_8} (\textbf{a}, \textbf{b} and \textbf{c}) show the positional distribution $\Psi_n(\xi)$ for $\beta=$ 0.171, 0.332 and 0.49 respectively. We observe similar trends as  in the positional distribution for ABP, where $\Psi(\xi)$ is very flat for an extended region close to the origin. For small $\beta$, $\xi=0$ is the peak and as $\beta$ increases, the peak shifts towards $\xi=1$. The figures in the bottom panel in Fig.~\ref{fig:Psi_chi_dist_RTP_dif_beta_for_n_8} show the corresponding angular distribution $f(\chi|\xi)$. Although the power $n$ is higher here, the angular distributions, which are unimodal and Gaussian for lower values of $\beta$ becomes flat as $\beta$ increases. Unlike the corresponding results for ABP, however, we do not observe the bimodal peaks clearly. 

The general features of the positional distribution is not very different for RTP, when compared to ABP. We see in Fig.~\ref{fig:xi_max_vs_beta_RTP_for_n_2_n8} that the shape transition appears continuous for $n=2$ and discontinuous for $n>2$. For $n=2$, the theoretical prediction (black dotted line) is close to the simulation data (Fig.~\ref{fig:xi_max_vs_beta_RTP_for_n_2_n8}\textbf{a}). For $n=8$, the data in Fig.~\ref{fig:xi_max_vs_beta_RTP_for_n_2_n8}\textbf{b} shows discontinuous transition with a critical value $\beta_c (8)\simeq 0.33$. 


\section{Conclusions and Discussion}
\label{conclusions}
In this work, we have explored the nonequilibrium stationary states of two types of self-propelling particles, active Brownian particle (ABP) and Run-and-Tumble particle (RTP) in general power-law potentials. Previous theoretical works in this direction have mostly been restricted to harmonic potentials and hard-wall confinements. Our work nicely interpolates between these two extremes and brings to light a number of interesting observations. 

To solve the Fokker-Planck equation (FPE), we essentially followed the same method as in our earlier work~\cite{nakul2023stationary}, in which the authors studied the stationary states of ABP in a harmonic potential. By integrating out the orientation angle, the FPE was reduced to a first order ordinary differential equation in the radial coordinate, which was solved explicitly under the assumption that the orientation angle is a Gaussian variable. In spite of the assumption, the resulting mathematical solution is found to be successful in reproducing the numerical simulation data. 

A remarkable finding we made during our work is that the shape transition in the positional probability density changes its nature as the power of the potential energy function is changed. While the transition is continuous for the harmonic potential ($n=2$), it becomes discontinuous for $n > 2$. Our theory successfully manages to predict the critical point of the transition in terms of a dimensionless activity parameter, as well as the gap parameter for the discontinuous transition. Our predictions are found to be in excellent agreement with numerical simulation results, the Gaussian approximation for the angle notwithstanding. 

Prior to our work, there have been experimental~\cite{dauchot2019dynamics} and theoretical~\cite{nakul2023stationary} papers which highlighted the existence of tangentially oriented orbiting states of ABP in a harmonic potential, in the limit of strong activity (in contrast to the radially oriented climbing states in the opposite limit). We have made substantial analytical progress in our understanding of such orientational states, and the transition between them. Our theoretical arguments identify this transition as a bifurcation, where the two mirror-image solutions correspond to clockwise and anti-clockwise orientations of the ABP. Here too, the analytical predictions manage to reproduce the numerical results quite well, although some observed features such as the horizontal shift in the origin of the bifurcation fork could not be explained within our theory. We hope to address these issues in future. 

\begin{acknowledgments}
We acknowledge useful discussions with V. Vasisht. A.S acknowledges the P.G. Senapathy Centre for Computing Resources, IIT Madras for computational facilities, the Soft Active Fluids Laboratory in the Department of Chemical Engineering, IIT Madras for financial support and the Department of Physics, IIT Palakkad for hospitality. 
\end{acknowledgments}

\appendix
\section{Details of the derivation of the terms in  Eq.~\ref{FPE_in_d_dimension}\label{Integration_I_1_and_I_2}}

Define the integrals 
\begin{equation}
I_1 = \int (\boldsymbol{\nabla} \cdot \mathbf{J}_t) \, (u_0 \hat{\mathbf{u}} - \frac{\epsilon_0}{\gamma_t R} \, \xi^{n-1} \hat{\mathbf{r}}) \, d \hat{\mathbf{u}}
\label{eq:I1}
\end{equation}
and 
\begin{equation}
I_2 = D_r \int (\nabla^2_{\hat{\mathbf{u}}} P) \, (u_0 \hat{\mathbf{u}} - \frac{\epsilon_0}{\gamma_t R} \, \xi^{n-1} \hat{\mathbf{r}}) \, d\hat{\mathbf{u}}
\label{eq:I2}
\end{equation}
Using Eq.~\ref{eq3}, Eq.~\ref{eq:I1} is simplified as 
\begin{equation} \label{I1_calculation}
I_1  = u_0 \bigg\{\int (\boldsymbol{\nabla} \cdot \mathbf{J}_t) \, \hat{\mathbf{u}} \, d\hat{\mathbf{u}} - \, \xi^{n-1} \hat{\mathbf{r}} \int (\boldsymbol{\nabla} \cdot \mathbf{J}_t)d\hat{\mathbf{u}}\bigg\}, 
\end{equation}
of which, the second integral vanishes identically, since $\int {\bf J}_t \,d{\hat {\bf u}}=0$. After substituting the explicit expression for ${\bf J}_t$, the first integral becomes 

\begin{equation}
\begin{split}
I_1 = &-D_t u_0 \int (\nabla^2 P)\, \hat{\mathbf{u}}\, d \hat{\mathbf{u}}\, + u_0^2 \bigg\{\int \hat{\mathbf{u}} \, (\hat{\mathbf{u}} \cdot \boldsymbol{\nabla} P) \, d\hat{\mathbf{u}}\\
&- \int [\boldsymbol{\nabla} \cdot (\xi^{n-1} \hat{\mathbf{r}} P)] \,\hat{\mathbf{u}} \, d \hat{\mathbf{u}}\bigg\}.
\end{split}
\end{equation}
We ignore the first term in the limit $D_t^{\prime}\to 0$, whence, after expanding the divergence, we find 
\begin{equation}
    \begin{split}
     I_1 \simeq & \,\, u_0^2 \bigg\{\int \hat{\mathbf{u}} (\hat{\mathbf{u}} \cdot \boldsymbol{\nabla} P) \, d \hat{\mathbf{u}} \, 
  - \int P \,\hat{\mathbf{u}} \, \boldsymbol{\nabla} \cdot (\xi^{n-1} \hat{\mathbf{r}})\, d \hat{\mathbf{u}} \\
 & - \, \xi^{n-1} \int (\hat{\mathbf{r}} \cdot \boldsymbol{\nabla}P)\, \hat{\mathbf{u}} \, d \hat{\mathbf{u}}\bigg\},
    \end{split}
\end{equation}

After using the following identities
\begin{equation}
    \boldsymbol{\nabla} \cdot (\xi^{n-1} \hat{\mathbf{r}}) = \frac{(n-2+d)}{R} \xi^{n-2}
\end{equation}
and
\begin{equation}
   \int (\hat{\mathbf{r}} \cdot \boldsymbol{\nabla} P) \, \hat{\mathbf{u}} \, d \hat{\mathbf{u}} \\
     = \frac{1}{R^2} \, \xi^{n-3} \,{\mathbf{r}}\, ({\mathbf{r}} \cdot \boldsymbol{\nabla} \Phi) + \frac{(n-1)}{R^2}\, \Phi\, \xi^{n-3} \,{\mathbf{r}}, 
\end{equation}

we find
\begin{equation}\label{final_I1}
\begin{split}
   I_1 = & \,u_0^2 \bigg\{\int \hat{\mathbf{u}} (\hat{\mathbf{u}} \cdot \boldsymbol{\nabla} P) \, d \hat{\mathbf{u}}- \frac{1}{R^2} \, (2n -3 +d)\, \Phi\, \xi^{2n-4} \, \mathbf{r}\,\\
   & - \frac{1}{R^2} \xi^{2n-4} \mathbf{r} \, (\mathbf{r} \cdot \boldsymbol{\nabla} \Phi)\bigg\}.
\end{split}
\end{equation}

Likewise, using Eq.~\ref{eq3}, we find  
\begin{equation}
    I_2 = u_0 D_r \bigg\{\int \hat{\mathbf{u}} \,(\nabla_{\hat{\mathbf{u}}}^2 P) \,  d\hat{\mathbf{u}} - \, \xi^{n-1} \hat{\mathbf{r}} \int \nabla_{\hat{\mathbf{u}}}^2 P \, d \hat{\mathbf{u}}\bigg\},
\end{equation}
of which, the second term vanishes identically. Further, by using the following general result~\cite{celani2010bacterial}, we find 
\begin{equation}
    \int  \hat{\mathbf{u}} \,\nabla_{\hat{\mathbf{u}}}^2 P \, d \hat{\mathbf{u}} = (1-d) \, \Phi \,\Bar{\hat{\mathbf{u}}}({r}),
\end{equation}
and therefore
\begin{equation}\label{final_I2}
    I_2 = u_0\, D_r (1-d) \, \Phi \,\xi^{n-1} \hat{\mathbf{r}}.
\end{equation}
After multiplying Eq.~\ref{eq:SFPE} by $\mathbf{v}$, and then integrating the equation over $\hat{\mathbf{u}}$ on both sides, we find the following expression
\begin{equation}
    I_1 - I_2 = 0,
\end{equation}
where $I_1$ and $I_2$ are defined in Eq.~\ref{eq:I1} and Eq.~\ref{eq:I2} respectively and with substituting the values of $I_1$ and $I_2$ from Eq.~\ref{final_I1} and Eq.~\ref{final_I2}, we have the resultant form of the above equation as,
\begin{equation}\label{I_1_and_I_2_final}
    \begin{split}
        &  u_0^2 \int \hat{\mathbf{u}} (\hat{\mathbf{u}} \cdot \boldsymbol{\nabla} P) \, d \hat{\mathbf{u}} \, - \frac{u_0^2}{R^{2}} \, (2n -3 +d)\, \Phi\, \xi^{2n-4} \, \mathbf{r}\, \\
        & - \frac{u_0^2}{R^{2}}\, \xi^{2n-4} \,\mathbf{r} \, (\mathbf{r} \cdot \boldsymbol{\nabla} \Phi)\\
        & - u_0\, D_r (1-d) \, \Phi \, \frac{1}{R}\, \xi^{n-2} \,{\mathbf{r}} =0.
    \end{split}
\end{equation}

\section{Calculation of the term $\hat{\mathbf{r}} \cdot \int \hat{\mathbf{u}} \, (\hat{\mathbf{u}} \cdot \boldsymbol{\nabla}_{\xi} \mathcal{P}) \, d \theta$}\label{calculation_of_u_u_gradP_int}
Given the axial symmetry of the potential $U(r)$, the stationary state probability distribution of the ABP 
$\mathcal{P}({\xi},\phi,\theta)$ should be invariant with respect to an in-plane coordinate rotation. Hence, the orientation angle $\theta$ enters the distribution only through the combination $\theta-\phi\equiv \chi$ (see Fig.~\ref{fig:schematic_abp_rtp}). For the same reason, the distribution cannot explicitly depend on $\phi$ either. 

By making use of the identities $\boldsymbol{\nabla}_{\xi}  \equiv \bigl(\hat{\mathbf{r}} \frac{\partial}{\partial \xi} + \frac{\hat{\boldsymbol{\phi}}}{\xi} \frac{\partial}{\partial \phi}  \bigr)$ and $\frac{\partial \chi}{\partial \phi} = -1$, we find 
\begin{equation}
    \hat{\mathbf{u}} \cdot \boldsymbol{\nabla}_{\xi} \mathcal{P} = \Bigl( \cos \chi \frac{\partial \mathcal{P}}{\partial \xi} - \frac{\sin \chi}{\xi} \, \frac{\partial \mathcal{P}}{\partial \chi} \Bigr)
    \label{eq:B1}
\end{equation}
where,
\begin{equation}
  \hat{\mathbf{u}} = \hat{\mathbf{r}} \cos \chi + \hat{\boldsymbol{\phi}} \sin \chi.  
  \label{eq:B2}
\end{equation}
The integration variable may now be replaced with $\chi$: 
\begin{equation}\label{radial_part_int_uudot_gradP}
\begin{split}
  &\hat{\mathbf{r}} \cdot \int \hat{\mathbf{u}} \, (\hat{\mathbf{u}} \cdot \boldsymbol{\nabla}_{\xi} \mathcal{P}) \, d\theta\\
  & = \int \cos^2\chi \,\frac{\partial \mathcal{P}}{\partial \xi} \, d\chi - \int \frac{\cos \chi \, \sin \chi}{\xi} \, \frac{\partial \mathcal{P}}{\partial \chi} \, d\chi.
\end{split}
\end{equation}
Define the conditional probability density $f(\chi|\xi)$ through the Bayesian relation ${\mathcal P}(\xi,\chi)=\Psi(\xi)f(\chi|\xi)$. Substitute this relation in Eq.~\ref{radial_part_int_uudot_gradP} to find 
\begin{equation}
\begin{split}
  & \hat{\mathbf{r}} \cdot \int \hat{\mathbf{u}} \, (\hat{\mathbf{u}} \cdot \boldsymbol{\nabla}_{\xi} \mathcal{P}) \, d\theta\\
  &= \frac{\partial}{\partial \xi} \Big[ \big< \cos^2\chi \big> \Psi(\xi) \Big] + \frac{\Psi(\xi)}{\xi} \Big[ 2\big< \cos^2 \chi \big> - 1 \Big]
\end{split}
\label{eqb4}
\end{equation}
where 
\begin{equation}
    \langle\cos^m\chi \rangle = \int \cos^m \chi \, f(\chi | \xi) \,d\chi. ~~(m=1,2,...)
\end{equation}

Define variance of $\cos \chi$: $\sigma^2_{\cos \chi} \equiv \sigma^2 = \big< \cos^2 \chi \big> - \big< \cos \chi \big>^2$. Then in terms of $\sigma^2$, Eq.~\ref{eqb4} becomes 
\begin{equation}\label{eq:B_4}
\begin{split}
  &\hat{\mathbf{r}} \cdot \int \hat{\mathbf{u}} \, (\hat{\mathbf{u}} \cdot \boldsymbol{\nabla}_{\xi} \mathcal{P}) \, d\theta\\
  &= \frac{\partial}{\partial \xi} \Big[ (\sigma^2 + \big< \cos\chi \big>^2) \Psi \Big] + \frac{\Psi}{\xi} \Big[ 2(\sigma^2 + \big< \cos \chi \big>^2) - 1 \Big].
\end{split}
\end{equation}
For an ABP in a power law potential, in the limit $D_t^{\prime}\to 0$, $\langle \cos\chi\rangle$ is given by Eq.~\ref{eq:g_of_chi}. Substitute the same in Eq.~\ref{eq:B_4} to find 

\begin{equation}\label{eq:B_8}
    \begin{split}
          &\hat{\mathbf{r}} \cdot \int \hat{\mathbf{u}} \, (\hat{\mathbf{u}} \cdot \boldsymbol{\nabla}_{\xi} \mathcal{P}) \, d \hat{\mathbf{u}} \\
        &= \Big[\sigma^2 + \, \xi^{2n-2}  \Big]\, \partial_{\xi}\Psi + \Psi \Big[\partial_{\xi}\sigma^2 + \frac{2\sigma^2}{\xi} - \frac{1}{\xi} + 2n \, \xi^{2n-3} \Big].
    \end{split}
\end{equation}

\section{Details of the solution of Eq.~\ref{diffential_eq_FP}}\label{simplification_int_dPsi_by_Psi}
Direct integration of Eq.~\ref{diffential_eq_FP} yields

\begin{equation}
\ln \Psi = \tilde{I}_1 + \tilde{I}_2
\label{eqc1}
\end{equation}
where
\begin{equation}
    \begin{split}
      \tilde{I}_1 =  & 2(2n-1)\int \frac{\, \xi^{2n-3} }{\Bigl[ 1 - \, \xi^{2(n-1)}  \Bigr]} d\xi\\
        & = \ln \Big[1 - \, \xi^{2(n-1)}  \Big]^{-\frac{(2n-1)}{(n-1)}} + {\rm const}, 
    \end{split}
    \label{eqc2}
\end{equation}
and 
\begin{equation}
    \begin{split}
       \tilde{I}_2 = &  -\frac{2}{\beta} \int \frac{\xi^{n-1} d\xi}{\Bigl[ 1 - \, \xi^{2(n-1)}  \Bigr]^2} \\
        &  = -\frac{2}{m\beta}\int \frac{z^{\frac{1}{m}} \,dz}{(1-z^2)^2},
        \label{eqc3}
    \end{split}
\end{equation}
where $m=n-1$ and $z=\xi^{n-1}$. The integral in Eq.~\ref{eqc3} can be expressed as the sum of four integrals: 
\begin{equation}
\label{eqc4}
\begin{split}
&\int \frac{z^{\frac{1}{m}} \,dz}{(1-z^2)^2}\\
&= \frac{1}{4} \int z^{\frac{1}{m}} \Big[\frac{1}{(1-z)^2} + \frac{1}{(1+z)^2} + \frac{1}{(1-z)} + \frac{1}{(1+z)}  \Big]\, dz
\end{split}
\end{equation}
Each integral appearing in the r.h.s of the above expression can be expressed in terms of Gauss hypergeometric functions. Consider the first integral, for example. We start with the obvious equality

\begin{equation}
I_A\equiv \int \frac{z^{\frac{1}{m}} dz}{(1-z)^2} = \int^z_0 \frac{t^{\frac{1}{m}}dt}{(1-t)^2}
\end{equation}
up to an additive constant. Now, define $s=t/z$ so that 

\begin{equation}
I_A=\int^1_0 \frac{s^{\frac{1}{m}}z^{\frac{1}{m} +1}ds}{(1-sz)^2}
\end{equation}
Using the definition of the Gauss Hypergeomertic function (see \cite{NIST:DLMF}), it follows that  

\begin{equation}
I_A= \frac{z^{(1+\frac{1}{m})} \, {}_2F_1\bigl(2,1+\frac{1}{m};2+\frac{1}{m};z\bigr)}{(1+\frac{1}{m})} \, +{\rm const}
\end{equation}
Similarly, the following relations can be derived:
\begin{equation}
    \begin{split}
     &\int \frac{z^{\frac{1}{m}} dz}{(1-z)^2}\\
     & = \frac{z^{(1+\frac{1}{m})} \, {}_2F_1\bigl(2,1+\frac{1}{m};2+\frac{1}{m};-z\bigr)}{(1+\frac{1}{m})} \, +{\rm const}   
    \end{split}
\end{equation}
\vspace{-0.6cm}
\begin{equation}
    \begin{split}
        &\int \frac{z^{\frac{1}{m}} dz}{(1-z)}\\
        & = \frac{z^{(1+\frac{1}{m})} \, {}_2F_1\bigl(1,1+\frac{1}{m};2+\frac{1}{m};z\bigr)}{(1+\frac{1}{m})} \, +{\rm const}
    \end{split}
\end{equation}
\vspace{-0.6cm}
\begin{equation}
    \begin{split}
        &\int \frac{z^{\frac{1}{m}} dz}{(1+z)}\\
        & = \frac{z^{(1+\frac{1}{m})} \, {}_2F_1\bigl(1,1+\frac{1}{m};2+\frac{1}{m};-z\bigr)}{(1+\frac{1}{m})} \, +{\rm const}
    \end{split}
\end{equation}

By using the standard series expansions for the hypergeometric functions, we find 
\begin{equation}\label{series_expansion}
    \begin{split}
        &  {}_2F_1\big(2,1+\frac{1}{m};2+\frac{1}{m};z \big) +\, {}_2F_1\big(2,1+\frac{1}{m};2+\frac{1}{m};-z \big) \\
        &   +\,  {}_2F_1\big(1,1+\frac{1}{m};2+\frac{1}{m};z \big) +\, {}_2F_1\big(1,1+\frac{1}{m};2+\frac{1}{m};-z \big) \\
        &   = 4 + \Big(1+\frac{1}{m} \Big) \Big[ \sum^{\infty}_{s=1} \frac{(s+2)}{(s+1+\frac{1}{m})} \big\{ z^s + (-z)^s \big\} \Big] \\       
    \end{split}
\end{equation}
\\
From Eq.~\ref{eqc3}, Eq.~\ref{eqc4} and Eq.~\ref{series_expansion}, we find, in terms of $\xi=z^{1/(n-1)}$, 

\begin{equation}
    \tilde{I}_2 = - \frac{2\xi^n}{n\beta} \Biggl\{1+ \sum^{\infty}_{s=1} \frac{(s+1)(\frac{n}{n-1})}{(2s+\frac{n}{n-1})} \xi^{2s(n-1)} \Biggr\}+{\rm const}. 
    \label{eqc12}
\end{equation}

Combining the expressions for $\tilde{I}_1$ 
(Eq.~\ref{eqc2}) and $\tilde{I}_2$ (Eq.~\ref{eqc12}), we get the final expression for the function $\Psi_n(\xi)$ in Eq.~\ref{eqc1}: 
\begin{equation}\label{appendix_expression_of_psi}
    \begin{split}
        &  \Psi_n(\xi) = C_n \Bigl[1 - \xi^{2(n-1)} \Bigr]^{-\frac{(2n-1)}{(n-1)}} \cdot\\ 
        & \exp \Bigl[-\frac{2\xi^n}{n \beta} \Bigl\{ 1 + \sum_{s=1}^{\infty} \frac{(s+1)(\frac{n}{n-1})}{(2s + \frac{n}{n-1})} \xi^{2s(n-1)} \Bigr\}   \Bigr]
    \end{split}
\end{equation}
where $C_n$ is a normalisation constant. 
\section{Approximate $\Psi_n(\xi)$ for large $n$}\label{aprrox_Psi_for_large_n}
Let us rewrite Eq.~\ref{appendix_expression_of_psi} in the form 
\begin{equation}\label{appendix_expression_of_psi_large_n}
    \begin{split}
        &  \Psi_n(\xi) = C_n\Bigl[1 -  \xi^{2(n-1)} \Bigr]^{-\frac{(2n-1)}{(n-1)}} \exp \Bigl[-\frac{2 \xi^n}{n \beta} \Bigl\{ 1 + \mathcal{F}_n(\xi) \Bigr\}   \Bigr]
    \end{split}
\end{equation}
where, in terms of $z_n=\xi^{2(n-1)}$ and $\alpha_n=n/(2n-2)$, the function $\mathcal{F}_n(\xi)$ is given by the expression
\begin{equation}
    \begin{split}
       & \mathcal{F}_n(\xi) = \alpha_n \Bigl[ \frac{z_n}{1-z_n} +(1-\alpha_n)\sum^{\infty}_{s=1} \frac{z_n^s}{s+\alpha_n}  \Bigr].
       \label{eq:FNXI}
    \end{split}
\end{equation}
The infinite series in Eq.~\ref{eq:FNXI} can be expressed in an integral form as in the following equation: 
\begin{equation}
    \begin{split}
        &  \sum^{\infty}_{s=1} \frac{z_n^s} {s+\alpha_n}=
        \mathcal{Q}_n(\xi)- \frac{1}{\alpha_n}, 
    \end{split}
\end{equation}
where 
\begin{equation}
\mathcal{Q}_n(\xi) = \int^{\infty}_0 \frac{\exp(-\alpha_n y) \, dy}{\bigl(1- z_n \exp(-y)\bigr)}, 
\label{eq:QNXI}
\end{equation}
using which we can express Eq.~\ref{eq:FNXI} as 
\begin{equation}
    \begin{split}
        & \mathcal{F}_n(\xi) 
         =\frac{\alpha_n \xi^{2(n-1)}}{1-\xi^{2(n-1)}} - (1-\alpha_n) + \alpha_n(1-\alpha_n) \mathcal{Q}_n(\xi)
    \end{split}
\end{equation}
For large $n$, $\alpha_n\to 1/2$. It can then be shown that 
$\mathcal{Q}_n(\xi)\to \mathcal{Q}_{\infty}(\xi)$ where 
\begin{equation}
    \begin{split}
        & \mathcal{Q}_{\infty}(\xi)=\\
        & = \frac{2}{\xi^{n-1}} \tanh^{-1}(\xi^{n-1}).
    \end{split}
\end{equation}
It follows that, for large $n$, we may approximate 
    \begin{eqnarray}
    \label{closed_form_expression_large_n}
      &  \Psi(\xi) 
      \simeq C_n \Bigl[1 -  \xi^{2(n-1)} \Bigr]^{-\frac{(2n-1)}{(n-1)}}\exp \Bigl[-\frac{\xi^n }{(n-1) \beta [1-\xi^{2(n-1)}]}\Bigr]\times \nonumber\\
     &  \exp \Bigl[-\frac{(n-2) \xi}{(n-1)^2 \beta} \tanh^{-1}(\xi^{n-1})\Bigr].
\end{eqnarray}


\clearpage
\bibliography{apssamp}

\end{document}